\documentclass[12pt]{iopart}
\usepackage{cite}
\usepackage[usenames,dvipsnames]{xcolor}
\definecolor{ao(english)}{rgb}{0.0, 0.5, 0.0}
\usepackage[breaklinks,colorlinks=true,linkcolor=blue,citecolor=ao(english)]{hyperref}
\usepackage[title,titletoc]{appendix}
\usepackage[english]{babel}
\usepackage[utf8]{inputenc}
\usepackage[OT2,T1]{fontenc}
\usepackage{graphicx}
\expandafter\let\csname equation*\endcsname\relax
\expandafter\let\csname endequation*\endcsname\relax
\usepackage{amssymb,amsfonts,amsthm}
\usepackage{newtx}
\usepackage{iopams}
\usepackage{bm,bbm}
\usepackage{tikz,stackrel}
\usepackage{silence} 
\usepackage[]{mdframed}

\newcommand{\bc}{\begin{center}}
\newcommand{\ec}{\end{center}}
\newcommand{\be}{\begin{equation}}
\newcommand{\ee}{\end{equation}}
\newcommand{\ba}{\begin{eqnarray}}
\newcommand{\ea}{\end{eqnarray}}
\def\bs{\begin{subequations}}
\def\es{\end{subequations}}
\renewcommand{\leq}{\leqslant}

\def\GWB{GWB\ }
\def\GWBn{GWB}

\newtheorem{cor}{Corollary}

\newcommand{\Eq}[1]{(\ref{#1})}

\def\cob{\color{blue}}

\newcommand{\au}[2]{#2 #1}
\newcommand{\ua}[2]{#1 #2}
\newcommand{\book}[5]{\emph{#1} (#3: #2)}

\newcommand{\oarX}[1]{\href{http://arxiv.org/abs/#1}{{\ttfamily\cob arXiv:#1}}}
\newcommand{\arX}[1]{\href{http://arxiv.org/abs/#1}{{\ttfamily\cob arXiv:#1}}}
\newcommand{\doin}[6]{\href{http://dx.doi.org/#1}{{\cob {\it #2} #3 {\bf #4} #5}}}
\newcommand{\doinn}[5]{\href{http://dx.doi.org/#1}{{\cob {\it #2} {\bf #3} #4}}}
\newcommand{\doij}[5]{\href{http://dx.doi.org/#1}{{\cob {\it #2} #3(#5)#4}}}

\newcommand{\procsinm}[5]{\emph{#1} ed #2 (#4: #3)}
\newcommand{\tia}[1]{#1}

\def\a{\alpha}

\def\g{\gamma}
\def\la{\lambda}

\def\e{\epsilon}

\def\Om{\Omega}
\def\om{\omega}

\def\s{\sigma}

\def\cA{\mathcal{A}}

\def\cF{\mathcal{F}}

\def\cN{\mathcal{N}}

\def\cP{\mathcal{P}}

\def\cT{\mathcal{T}}

\def\cob{\color{blue}}

\newcommand{\Pl}{{\textrm{\tiny Pl}}}

\newcommand{\Mpl}{M_\Pl}
\def\rme{{\rm e}}
\def\rmd{{\rm d}}
\def\rmi{{\rm i}}

\newcommand{\nn}{\nonumber\\}

\makeatletter
\long\def\@makefntext#1{\parindent 1em\noindent 
 \makebox[1em][l]{\footnotesize\rm$\m@th{^\arabic{footnote}}$}%
 \footnotesize\rm #1}
\def\@makefnmark{\hbox{$^{\arabic{footnote}}\m@th$}}
\def\@thefnmark{\arabic{footnote}}
\makeatother

\begin{document}

\title{Log-periodic gravitational-wave background beyond Einstein gravity}

\author{Gianluca Calcagni}
\ead{g.calcagni@csic.es}
\address{Instituto de Estructura de la Materia, CSIC, Serrano 121, 28006 Madrid, Spain}

\author{Sachiko Kuroyanagi}
\ead{s.kuroyanagi@csic.es}
\address{Instituto de F\'isica Te\'orica UAM-CSIC, Universidad Aut\'onoma de Madrid, Cantoblanco, 28049 Madrid, Spain}
\address{Department of Physics, Nagoya University, Furocho, Chikusa, Nagoya, 464-8602, Japan}

\begin{abstract}
Periodic patterns of logarithmic oscillations can arise in primordial curvature perturbation spectra and in the associated gravitational-wave background via different mechanisms. We show that, in the presence of log oscillations, the spectral shape of the stochastic background has a unique parametrization independent of its physical origin. We also show that this log-periodic modulation can be generated in any scenario beyond Einstein gravity endowed with an approximate discrete scale invariance, a symmetry typical of deterministic fractal spacetimes that could emerge in quantum gravity under certain conditions. We discuss how a log-oscillatory spectral shape arises from concrete inflationary models beyond Einstein gravity and the prospects for detection in Einstein Telescope and other next-generation gravitational-wave observatories. We find that these instruments will be sensitive to log-periodic features if the detection is made with a high signal-to-noise ratio (SNR) and that the error scales as $1/{\rm SNR}$.
\end{abstract}
\begin{indented}
\item[]August 11, 2023\hfill ET-0274A-23\qquad IFT-UAM/CSIC-23-105
\end{indented}
\vspace{10pt}


\centerline{2023 \doinn{10.1088/1361-6382/ad1123}{Class.\ Quantum Grav.}{41}{015031}{2023} [\arX{2308.05904}]}


\section{Introduction}

The discovery of gravitational waves (GWs) in 2015 \cite{Abbott:2016blz,TheLIGOScientific:2016src} heralded a new era of observations that promise to share light on a number of questions in astrophysics and stellar evolution, cosmology and gravitation. While the confirmation of Einstein's theory of general relativity and the verification of the main models of black holes and binary mergers remain the core science pursued by large collaborations such as LIGO-Virgo-KAGRA (LVK), LISA, Einstein Telescope (ET) and Cosmic Explorer (CE), there is also latitude for exploring and constraining alternative scenarios of gravity and of the physics of compact objects. In particular, this acknowledged capability of GW astronomy with present- and next-generation instruments has been driving efforts to understand what GWs can say about the fundamental nature of gravity and classical or quantum departures from Einstein's theory \cite{Mirshekari:2011yq,Canizares:2012is,Ellis:2016rrr,Yunes:2016jcc,Arzano:2016twc,Calcagni:2019kzo,Belgacem:2019pkk,Calcagni:2019ngc,Belgacem:2020pdz,Barausse:2020rsu,Calcagni:2020tvw,Addazi:2021xuf,LISACosmologyWorkingGroup:2022wjo,LISACosmologyWorkingGroup:2022jok,LISA:2022kgy}.

One of the main protagonists in this quest is the stochastic gravitational-wave background (\GWBn), the random superposition of GWs from different sources \cite{Christensen:2018iqi,Renzini:2022alw,vanRemortel:2022fkb}. Models of \GWB can usually be divided into two, those of relatively late astrophysical origin and those of primordial origin. Recently, a \GWB possibly consistent with a population of supermassive black-hole binaries has been indicated at frequencies $f\sim 10^{-9}\!-\!10^{-8}\,{\rm Hz}$ \cite{NANOGrav:2023gor,Antoniadis:2023ott,Reardon:2023gzh,Xu:2023wog}, while a \GWB relic of early-universe physics, expected at higher frequencies, remains elusive. In the latter case, an example of seeds is inflationary tensor perturbations. All early-universe models predict a certain spectral shape $\Om_{\textsc{gw}}(f)$, i.e., a \GWB amplitude as a function of the observed frequency $f$, which can be catalogued into templates. The simplest spectral shape is a single power law $\Om_{\textsc{gw}}(f)\sim f^{n_{\rm t}}$, corresponding to a featureless line with tilt $n_{\rm t}$ in logarithmic scale. However, in some cases the spectral shape can be non-monotonic or even have features. The type of features depends on the underlying model. Along the above lines, one can pose the following questions: Are there features in the \GWB that bear the imprint of fundamental physics beyond Einstein gravity? Can we observe them with present or future GW interferometers?

In this paper, we examine known and new theoretical models generating a \GWB spectral shape modulated by a specific type of features, logarithmic oscillations:
\ba
\hspace{-.8cm} \Om_{\textsc{gw}}(f) &=& \bar\Om_{\textsc{gw}}(f)\left[1+\sum_{l=1}^{l_{\rm max}} F_l(f)\right],\label{master1}\\
\hspace{-.8cm} F_l(f) &=& A^{\rm c}_l\cos\left(l\om\ln\frac{f}{f_{*}}\right)+A^{\rm s}_l\sin\left(l\om\ln\frac{f}{f_{*}}\right),\label{master2}
\ea
where in \Eq{master1} $f$ is the observed frequency of the GW signal, $\bar\Om_{\textsc{gw}}(f)$ is the envelope or main spectral shape along which the modulation takes place and $l$ labels the harmonics, up to some number $l_{\rm max}$ determined by the model. In \Eq{master2}, $f_{*}$ is the frequency scale corresponding to some characteristic momentum or length scale, $A^{\rm c}_l$ and $A^{\rm s}_l$ are constant amplitudes associated with, respectively, the cosine (c) and sine (s) term and, finally, $\omega$ is the frequency of the log oscillations, a theoretical or phenomenological parameter not to be confused with the observational parameter $f$.

Our main theoretical contribution (section \ref{maindsi} and \ref{appA}) is to show that the spectral shape \Eq{master1}--\Eq{master2} can arise in \emph{any} matter or gravitational theory enjoying discrete scale invariance (DSI) in a certain frequency (energy, length) range. To put it in practical terms, any \GWB whose plot displays log-periodic oscillations can be described by \Eq{master1}--\Eq{master2} for a non-trivial choice of the parameters therein. This result is model-independent but we discuss various generation mechanisms for a DSI that fall into two categories: scenarios where DSI is an emergent property due to matter (i.e., non-gravitational) fields and scenarios, which include models of the early universe coming from quantum gravity, where DSI is an intrinsic property of spacetime, either as a fundamental symmetry of spacetime geometry or an emergent one from certain choices of Fock vacua or from a modified dispersion relation for the graviton. 

For both categories, we give explicit examples of early-universe scenarios generating a log-periodic pattern in primordial power spectra and/or in the primordial \GWB as in \Eq{master1}--\Eq{master2}. Most of these models are known in the literature and we succinctly review them in section \ref{revi}. In some of the examples, the presence of a DSI was explicitly acknowledged in the original papers, while in others it was not, but all these models belong to the category of scenarios where DSI emerges from the dynamics of matter fields such as the inflaton, extra massive scalars or gauge $p$-forms. Note that, in most of these cases, modifications arise only in the scalar power spectrum while leaving the tensor spectrum unmodified or suppressed. Then, the \GWB may acquire a log modulation only when generated by second-order scalar perturbations.

In contrast, in section \ref{dsiqg} we consider scenarios of quantum gravity where DSI is a property of spacetime and we show how the tensor spectrum translates into a \GWB following the spectral shape \Eq{master1}--\Eq{master2}. After discussing DSI as a spacetime property in quantum gravity in section \ref{sec41} and reviewing the case of trans-Planckian inflation in section \ref{revi1}, in section \ref{mufr} we develop a DSInvariant model in the so-called multi-fractional theory with $q$-derivatives. Although this is not a consistent theory of quantum gravity in its simplest version \cite{Calcagni:2013qqa,Calcagni:2016azd}, it displays many of the geometric phenomena we expect in such a theory \cite{Calcagni:2016azd,Calcagni:2021ipd}. While neither the presence of log oscillations in the primordial spectra of this model \cite{Calcagni:2017via} nor the average trend of the \GWB \cite{Calcagni:2020tvw} are a novelty, here we derive the full, oscillatory \GWB spectral shape explicitly. Last, in section \ref{inter} we discuss the spectral shape \Eq{master1}--\Eq{master2} in relation to the observability of the log modulation by GW interferometers, with special focus on the Einstein Telescope \cite{Maggiore:2019uih} (similar for CE) but later commenting also on the cases of LISA and DECIGO. We find that log oscillations can be detected with one harmonic ($l_{\rm max}=1$) if the signal-to-noise ratio is ${\rm SNR}\gtrsim 100$ for the frequency parameter $\om\sim10$ and an amplitude $A_1\sim 0.1$. Even more interestingly, the error on the estimate of the amplitudes for higher-order harmonics does not increase with $l$, since the frequency of all these harmonics is an integer multiple of $\om$ and this parameter already gets constrained from the $l=1$ order. Conclusions are in section \ref{concl}.


\section{Discrete scale invariance in the \GWBn}\label{maindsi}

In this section, we argue that the spectral shape \Eq{master1}--\Eq{master2} arises in any cosmological model endowed with a fundamental or an emergent DSI.

The log-oscillatory part \Eq{master2} of the spectral shape is invariant under a discrete dilation symmetry in frequency space:
\be\label{eq:dsi}
F_l(\la_\om^m f)=F_l(f)\,,\qquad\la_\om\coloneqq \rme^{\frac{2\pi}{\om}}\,,\qquad m\in\mathbb{Z}\,.
\ee
In turn, this translates into a DSI in momentum space as well as a DSI in position space, $x^\mu\to \la_\om x^\mu$, all with the same scaling ratio $\la_\om$. The spectral shape \Eq{master1}--\Eq{master2} is not exactly DSInvariant, since the envelope $\bar\Om_{\textsc{gw}}(f)$ does not usually enjoy this symmetry:
\be
\bar\Om_{\textsc{gw}}(\la_\om^m f)\neq \bar\Om_{\textsc{gw}}(f)\,.
\ee
Nevertheless, any geometry or mechanism with a DSI and capable, at the same time, of generating a blue-tilted primordial \GWB can eventually end up with a spectral shape with the modulation factor \Eq{master2} in the range of future GW interferometers. In fact, in \ref{appA} we prove that
\begin{mdframed}
\emph{If a \GWB signal displays logarithmic oscillations at a certain frequency range, then it can always be written as \Eq{master1}--\Eq{master2} in that range, regardless of its physical origin.}
\end{mdframed}
This powerful but simple result is particularly useful in all cases where the \GWB is computed numerically from an integral equation but no explicit expression of the spectral shape is available in closed form.

On the other hand, if the modulation factor in \Eq{master1} is approximately but not exactly DSInvariant, the spectral shape \Eq{master1}--\Eq{master2} is not the most general one but it is still sufficient at the scales of GW interferometers. For instance, a generalization of the arguments of sine and cosine to powers of the logarithm would change the periodicity of the function by a non-integer factor and, thus, the corresponding scaling ratio. Setting schematically $f_*=1$,
\ba
\sum_s[\om\ln(\la_\om^m f)]^s&=&\sum_s(\om m\ln\la_\om+\om\ln f)^s\nn
&=&\sum_s(2\pi m+\om\ln f)^s\nn
&=&\sum_s\sum_{q=0}^s\binom{s}{q}(2\pi m)^{s-q}(\om\ln f)^q,\label{logn}
\ea
which would violate \Eq{eq:dsi}. However, even if higher-order terms in the sum break DSI, they can be neglected near the pivot scale $f\simeq f_*\simeq f_0$ (i.e., $f\approx 1$ in $f_*=1$ units, so that $\ln f\ll 1$) without affecting the precision reachable by GW interferometers. This would not be the case if the scalar and tensor spectra had the same quasi-DSI modulation at cosmic microwave background (CMB) scales, where measurements of temperature fluctuations are extremely precise and constraints on the amplitude of log oscillations may depend on the details of the spectral shape. Barring this scenario, for which we have only one example (section \ref{revi}), we can conclude that the application of the spectral shape \Eq{master1}--\Eq{master2} extends also to models where DSI is softly broken also in the modulation factor $\sum_lF_l$.

Having established that the spectral shape \Eq{master1}--\Eq{master2} captures any model with exact or very softly broken DSI, we turn to the question of what physical mechanisms would induce a DSI in cosmological models. Indeed, the usefulness of \Eq{master1}--\Eq{master2} would be very limited or even null if there were no models or theories displaying this type of symmetry.

In the case of a \GWB of primordial origin, the spectral shape \Eq{master1}--\Eq{master2} either arises as a second-order effect induced by the scalar sector or is directly generated within the tensor sector. In sections \ref{revi} and \ref{dsiqg}, we will discuss both classes of mechanisms.


\section{Models with matter-induced log oscillations}\label{revi}

In this section, we first review models where DSI in cosmological spectra emerges as a non-fundamental property from some mechanism which involves matter fields (scalars and gauge fields) and also includes the breaking of such symmetry. In general, second-order scalar perturbations can generate a \GWB \cite{Tom67,Matarrese:1992rp,Matarrese:1993zf,Matarrese:1997ay,Mollerach:2003nq} through a certain convolution of two copies of the primordial scalar spectrum $\cP_{\rm s}$ \cite{Ananda:2006af,Baumann:2007zm}. Moreover, if the scalar spectrum is log-oscillating, then the resulting GW spectral shape is of the form \Eq{master1}--\Eq{master2} with $l_{\rm max}=2$ harmonics and the amplitudes $A^{\rm c,s}_{1,2}$ depending on the other parameters \cite{Fumagalli:2021cel}. All the models listed in this section can produce a scalar-induced log-oscillatory \GWB via this mechanism, while their tensor spectrum is either standard or strongly suppressed. Therefore, these models fit CMB data and generate red-tilted spectra and, in particular, their primordial tensor sector remains directly unreachable by GW interferometers. Models predicting a large tensor-to-scalar ratio can nevertheless be constrained by present and future GW polarization observations.

In the class of inflationary models with resonant features, there are bumpy potentials or extra matter fields (scalar or vector) that do not drive inflation but participate to its dynamics. These dynamics usually induce sudden variations in the propagation speed of scalar perturbations which eventually result in log oscillations in the scalar power spectrum. The list below is, approximately, in chronological order of appearance in the literature.
\begin{itemize}
\item \emph{Potential with bumps}. Shaping the inflaton potential with one or more bumps \cite{Chen:2008wn} generates a log-oscillating scalar spectrum while leaving the tensor one unaltered. These characteristics could therefore translate to the \GWB only via induction by second-order scalar fluctuations. The three-point correlation function is also affected and this model, like others in this  list, produces non-Gaussianity \cite{Chen:2008wn}.
\item \emph{Sudden trajectory turns}. If the inflaton is the only light scalar among many others and if it experiences sudden turns in the trajectory in the multi-field potential \cite{Achucarro:2010jv,Achucarro:2010da} or other types of sharp features, then wiggles of logarithmic periodicity induced by an oscillating massive field may appear in the scalar primordial spectrum \cite{Chen:2011zf,Chen:2011tu,Shiu:2011qw,Gao:2012uq,Gao:2013ota,Noumi:2013cfa,Chen:2014joa,Chen:2014cwa,Braglia:2020taf}. In this group of models, the multi-field potential has a flat but bending `gorge' surrounded by steep walls. The inflaton slowly rolls at the flat bottom of the gorge, which is the minimum of the potential of spectator massive fields (the directions in the field space transverse to the trajectory). Due to one or more sharp features of the inflaton trajectory, the massive fields are suddenly displaced from the minimum and undergo quick oscillations around it. The resonance triggered after the initial response can produce logarithmic oscillations in a certain frequency range. In the two-field case (inflaton $\phi$ plus a massive scalar $\s$ with mass $m_\s$), if the background accelerates fast (extreme slow-roll regime) and for certain potentials $V(\phi,\s)=V_1(\phi)+V_2(\s)$, the scalar spectrum is modulated by sines and cosines of a logarithm \cite{Chen:2011zf,Chen:2011tu,Gao:2013ota,Chen:2014joa,Chen:2014cwa} and the resulting oscillations have a high frequency $2m_\s/H\gg 1$ (where $H=\dot a/a$ is the Hubble parameter) and, contrary to most of the other cases, zero average (no constant term 1 in the harmonic series).
\item \emph{Axion monodromy inflation}. In axion monodromy inflation, a cosmological scenario arising in flux compactification in string theory, the amplitude and frequency of the log oscillations are determined by the expectation values of moduli fields. The power spectra contain powers of the logarithm $\sim\cos[\Phi+\ln (k/k_0)+\ln^2 (k/k_0)+\dots]$ (where $\Phi$ is a phase), as in \Eq{logn} \cite{Flauger:2009ab,Flauger:2014ana} (see also the generalized scenario of \cite{Behbahani:2011it}). Higher-order terms can be neglected near the pivot scale $k\simeq k_0$, but only starting from the fourth power in the case of CMB frequencies. Therefore, DSI in momentum or frequency space is only approximate here and the single-harmonic spectral shape \Eq{cpt2} below (with $A^{\rm s}_1=0$) would be more suitable at higher (interferometer) frequencies. Recently, the sharp-turn scenario described above has been realized in axion monodromy inflation \cite{Bhattacharya:2022fze}. In any case, the main effect is confined to the scalar sector, while log oscillations in the tensor spectrum are strongly suppressed. Therefore, in this model only a scalar-induced \GWB could be observable as far as GW cosmology is concerned.
\item \emph{Fluxes in non-minimally coupled gravity}. DSI generated by fluxes is also a central ingredient of a different model outlined in \cite{Calcagni:2017via}, where one has a dynamical soft breaking of continuous scale invariance $x^\mu\to\la x^\mu$, $\la\in\mathbb{R}$, a sub-symmetry of conformal-invariant field theories. In this case, the scaling ratio $\la$ is the expectation value of a field. In a setting similar to flux compactification in low-energy string theory \cite{Giddings:2001yu}, one could consider a $p$-form $A_p$ on a compact subspace $\Gamma_{p+1}$, which produces a quantized flux $\int_{\Gamma_{p+1}}\rmd A_p= n q$, where $q\eqqcolon 2\pi/\om$ is the $p$-form magnetic charge defining a parameter $\om=2\pi/q$. An effective action $S[g_{\mu\nu},A_p]$ characterized by a conformally coupled metric $\exp(2\int \rmd A_p)\,g_{\mu\nu}$ is associated with a coordinate dilation with arbitrary scaling ratio $\lambda\sim \exp(\int \rmd A_p)$. However, upon quantizing the flux, one forces the system to a discretized conformal coupling $\la_\om^{2n}g_{\mu\nu}$, where $\la_\om$ is given in \Eq{eq:dsi}. Then, a DSI arises dynamically from fluxes. In order to obtain a superposition of harmonics with different frequencies, one can extend this mechanism to different $p$-forms with different charges $q$. Details of this scenario have not been studied yet.
\item \emph{Axion Chern--Simons gravity}. In this model motivated by string-theory compactifications, the inflaton is one of two axions and its potential is modulated by linear oscillations in the field \cite{Mavromatos:2022yql}. The resulting scalar spectrum, to which also the other axion contributes with its dynamics, is log-oscillatory.
\item \emph{Complex Lifshitz scalars}. Lifshitz scalars are such that the spatial-derivative operator in the kinetic terms is of higher order than the time-derivative one. In this class of models, one considers a complex Lifshitz scalar field $\phi$ with a space-dependent potential $V(\bm{x},\phi)$. The system undergoes a universal phase transition between a continuous and a discrete scale invariance where the DSI phase is characterized by an infinite set of bound states \cite{Brattan:2017yzx}. Just like the models described above, here DSI is generated dynamically from matter fields, without any underlying spacetime hierarchical structure. The inflationary spectra for this model have not been calculated yet and it is not even clear that a Lifshitz scalar can sustain inflation by itself, since solutions are manifestly space-dependent. A possibility is to embed $\phi$ in a multi-field configuration where another scalar plays the role of the inflation. Then, depending on how the symmetries of the Lifshitz scalar carry over to cosmological observables, the appearance of broken DSI in the primordial spectra is expected \cite{Calcagni:2017via}.
\end{itemize}


\section{Models with spacetime log oscillations}\label{dsiqg}

In this section, we consider scenarios of quantum gravity where DSI is an intrinsic property of spacetime rather than something emerging from matter fields or from the interaction between matter and gravity. We also derive explicitly the spectral shape \Eq{master1}--\Eq{master2} for a model living in a fractal spacetime.

Contrary to the models of section \ref{revi}, those of this section can produce both a scalar- and a tensor-induced primordial \GWBn. In both cases, the primordial (respectively, scalar and tensor) spectrum has the same symmetry properties as the spectral shape. To make this point, consider the direct-generation mechanism and a tensor spectrum with only one harmonic, $l_{\rm max}=1$:
\be\label{cpt2}
\hspace{-.7cm} \cP_{\rm t}(k)  = \bar\cP_{\rm t}(k)\left[1+A^{\rm c}_1 \cos\left(\om\ln\frac{k}{k_{*}}\right)+A^{\rm s}_1 \sin\left(\om\ln\frac{k}{k_{*}}\right)\right],
\ee
where $k$ is the comoving wave-number, $k_{*}$ is some characteristic momentum scale and the prefactor $\bar\cP_{\rm t}(k)$ depends on the model. To extract the spectral shape of the \GWB from the power spectrum \Eq{cpt2}, we recall the formula
\be\label{Omgw}
\Om_\textsc{gw}(f) =\frac{k^2}{12a_0^2H_0^2}\cP_{\rm t}(k)\,\cT^2(k, \tau_0)\Big|_{k=\frac{f}{2\pi}}\,,
\ee
where the subscript 0 indicates quantities evaluated at the present time (one can set $a_0=1$, while $H_0$ is the Hubble constant today) and $\cT(k, \tau_0)$ is the transfer function encoding how the primordial spectrum evolved after horizon crossing until today. Its specific form depends on the history of the universe \cite{Turner:1993vb,Kuroyanagi:2008ye} and it can be simplified in the inflationary scenario of instantaneous reheating \cite{Nakayama:2008wy,Kuroyanagi:2014nba}. Plugging \Eq{cpt2} into \Eq{Omgw}, we obtain \Eq{master1}--\Eq{master2} with $l_{\rm max}=1$ and the envelope
\be
\bar\Om_\textsc{gw}(f) =\frac{k^2}{12a_0^2H_0^2}\bar\cP_{\rm t}(k)\,\cT^2(k, \tau_0)\Big|_{k=\frac{f}{2\pi}}\,.\label{barOmgw}
\ee
This example illustrates the typical derivation of the \GWB spectral shape from the primordial tensor spectrum predicted by a cosmological model of the early universe and it is immediately extendable to the multi-harmonic case \Eq{master2}. In the presence of two or more harmonics, the pattern of peaks and troughs becomes more complicated and, in some case, it can lead to log-periodic spikes \cite{Calcagni:2017via}.


\subsection{DSI in quantum gravity}\label{sec41}

Discrete scale invariance is a landmark feature in chaotic, critical and hierarchical systems \cite{Sornette:1997pb} as well as in deterministic fractals where the scaling ratio $\la_\om$ fixes how copies of the fractal are replicated at different scales \cite{BGM1,DDSI,Bessis:1983nt,NLM,LvF}. In turn, DSI can be interpreted as the manifestation of a geometry with a complex dimension \cite{BGM1,LvF} whose imaginary part is identified with the parameter $\om$ \cite{Calcagni:2017via}. The value of $\om$ is model-dependent and, in spacetimes with a deterministic (multi-)fractal geometry, it is determined by the geometry itself. In particular, it depends both on the Hausdorff dimension of spacetime and on the number of copies $\cN$ produced by the iterative mapping process typical of fractals at each iteration \cite{Calcagni:2016edi}. To make a toy example of what we mean with these technical terms, the popular middle-third Cantor set is specified by an initial segment of unit length and by two maps ($\cN=2$) which rescale and double the number of copies of the segment at each iterative step.

Fractal spacetimes emerge in theories of quantum gravity as the result of, so to speak, the deformation of classical geometries by quantum mechanics \cite{Crane:1985ex,tHooft:1993dmi,Ambjorn:2005db,Lauscher:2005qz,Benedetti:2008gu} and they can be catalogued according to their geometric properties \cite{Calcagni:2019ngc,Carlip:2009kf,Calcagni:2009kc,Calcagni:2016xtk,Calcagni:2016azd,Carlip:2017eud,Mielczarek:2017cdp,Carlip:2019onx}. However, in most cases the `fractal' nature of these effective spacetimes is limited to a transition from an ultraviolet (UV, short-scale) geometry with exotic properties to an infrared (IR, large-scale) standard one, thus circumscribing the effects beyond Einstein gravity to very small scales or, in the most optimistic cases, requiring the cumulative enhancement of such effects via the propagation of GWs through cosmologically large distances \cite{Belgacem:2019pkk,Calcagni:2019ngc}. Such a possibility relies on the existence of a non-monotonic behaviour of the spectral dimension at intermediate scales, a feature present only in very few models. Hints that such a feature could happen in certain kinematical states in loop quantum gravity, spin foams and group field theory \cite{Calcagni:2014cza} have been confirmed only recently \cite{Jercher:2023rno}.
 
In contrast, DSI crosses the usual UV/IR divide characterizing other quantum-gravity effects and manifests itself at all scales. This happens because locally (accordingly to the frequency $\om$ and the dilation factor $\la_\om$ in \Eq{eq:dsi}) DSI replicates the same structure over and over again at arbitrarily small and arbitrarily large distances, even if the symmetry itself is explicitly broken in the IR. In this sense, DSI is not confined to Planckian scales or frequencies. The magnitude of the effect (amplitude of the log oscillations) may be small, but it is everywhere. However, it is still unclear where DSI makes its appearance in the landscape of theories of quantum gravity.

To summarize, we have two notions of `fractality:' one meant as a change of scaling of correlation functions across different scales and one meant as a property of self-similarity. The first is well established in quantum gravity and corresponds to what mathematicians call random fractals. The second is only beginning to pop up in quantum gravity and corresponds to deterministic fractals. There are reasons to believe that the search for DSI in quantum gravity will eventually yield positive results:
\begin{enumerate}
\item DSI or, more precisely, quasi-DSI as formulated in this paper does not violate the long-standing conjecture that a theory of quantum gravity should not possess exact global symmetries \cite{Banks:2010zn,Harlow:2018jwu,Harlow:2018tng}, either continuous or discrete, because it is broken explicitly: in the case of cosmological primordial spectra, by the envelope function. Thus, DSI is on the `safe side' regardless of whether the above conjecture holds true universally or only for certain quantum gravities such as those formulated within string theory or the AdS/CFT correspondence \cite{Harlow:2018jwu,Harlow:2018tng}.
\item DSI is a natural outcome of the most general factorized measure of spacetime with anomalous scaling \cite{Calcagni:2016xtk}. Moreover, random multi-fractals are generated from deterministic multi-fractals by a randomization of the scaling ratios of the latter. In the case of quantum gravity, this randomization process could be originated from the quantum fluctuations of the fundamental building blocks of spacetime, which would explain why anomalous scaling (the geometric behaviour typical of random multi-fractals) is seen in all theories where the gravitational field is quantized. Therefore, at least in some cases, this typical behaviour with a UV/IR divide may have a self-similar substratum. 
\item Connected with the last point, it is altogether possible that spacetimes effectively emerging from a fundamental geometric or pre-geometric framework display a deterministic fractal structure (hence a DSI) at certain scales \cite{Calcagni:2016azd,Carlip:2017eud}. So far, evidence is scant but building up. For example, in ongoing investigations in spin foams, there are not yet any signs for a complex dimension but this is mostly because one is still limiting explicit calculations and simulations to the semi-classical (large-spin) approximation of spin foams \cite{Steinhaus:2018aav} or due to other approximations and assumptions \cite{Jercher:2023rno}. Complexity is mostly avoided in this regime but it is indeed expected to arise from the measure in the spin foam \cite{Thu23}. 
\end{enumerate}
While theoretical efforts to find DSI are taking place in quantum gravity, it can be interesting to explore its phenomenological consequences. In section \ref{revi1}, we review phenomenological models of inflation with trans-Planckian physics, while in section \ref{mufr} we derive the spectral shape \Eq{master1}--\Eq{master2} from a cosmological model with an explicit DSInvariant modulation.


\subsection{Trans-Planckian inflation}\label{revi1}

Logarithmic oscillations arise in slow-roll inflationary models with a UV energy cut-off, a modified dispersion relation and/or a vacuum choice different from Bunch--Davies \cite{Martin:2000xs,Easther:2001fz,Easther:2001fi,Easther:2002xe,Martin:2003kp,Greene:2004fln,Chen:2010bka}. For example, a non-standard vacuum choice (induced or not by a modified dispersion relation) can lead to a normalization of the solution of the Mukhanov--Sasaki equation of cosmological perturbations with complex exponents $|\tau|^{\rmi\omega}=\exp(\rmi\om\ln|\tau|)$, where $\tau=\int\rmd t/a(t)$ is conformal time and $a(t)$ is the background scale factor in proper time. Reality of the solution linearly combines such exponents into trigonometric functions of a logarithm, $|\tau|^{\rmi\om}\pm|\tau|^{-\rmi\om}\sim\cos,\sin(\om\ln|\tau|)$, which eventually appear in the power spectrum after replacing conformal time $\tau=\tau(k)$ as a function of the comoving wave-number $k$. In general, these models generate a tensor power spectrum
\be\label{cpt}
\cP_{\rm t}(k)  = \bar\cP_{\rm t}(k)\left[1+A^{\rm c}_1(k) \cos B(k)+A^{\rm s}_1(k) \sin B(k)\right],
\ee
where
\be
\bar\cP_{\rm t}(k)=\frac{8}{\Mpl^2}\left[\frac{H(k)}{2\pi}\right]^2=\bar\cA_{\rm t}\left(\frac{k}{k_0}\right)^{n_{\rm t}},
\ee
is the standard spectrum in Einstein gravity to lowest slow-roll order, $\Mpl=(8\pi G)^{-1/2}$ is the reduced Planck mass and $H$ is the Hubble parameter evaluated at horizon crossing. In the second equality, we parametrized $\bar\cP_{\rm t}(k)$ in terms of the tensor amplitude $\bar\cA_{\rm t}=\bar\cP_{\rm t}(k_0)$ and the tensor spectral index $n_{\rm t}$. A similar expression holds also for the primordial scalar spectrum $\cP_{\rm s}(k)$. 

The functions $A^{\rm c,s}_1(k)$ and $B(k)$ are determined by the type of dispersion relation in the Mukhanov--Sasaki equation of the model and by the vacuum on which such equation is solved. Both these elements introduce characteristic scales that shape the profile and parameter dependence of the above functions. In particular, it is possible to get spectra with $B(k)\propto\ln k$ and exactly or approximately constant $A^{\rm c,s}_1$ from the choice of non-standard vacua with a UV cut-off \cite{Easther:2001fz,Easther:2001fi,Easther:2002xe,Martin:2003kp,Greene:2004fln}.\footnote{In other cases, the normalization factors are phases combined into a trigonometric function with linear oscillations, $B(k)\sim k$ \cite{Martin:2000xs,Martin:2003kp,Greene:2004fln,Danielsson:2002kx,Bozza:2003pr}. Oscillations are generated also when integrating out massive fields during inflation but they are not a general feature, since they depend on the choice of UV cut-off \cite{Jackson:2010cw}.} This leads to \Eq{cpt2}, where $\om$ is a model-dependent constant and $k_*=k_0$ is the pivot scale of the observation. As shown in section \ref{maindsi}, the spectral shape $\Om_{\textsc{gw}}(f)$ is indeed \Eq{master1}--\Eq{master2} with $l_{\rm max}=1$ in the case of a \GWB directly generated by the tensor spectrum ($\Om_{\textsc{gw}}$ has only as many harmonics as in $\cP_{\rm t}$, i.e., one), or $l_{\rm max}=2$ in the case it is generated by second-order scalar perturbations.

Let $\e=-\dot H/H^2$ be the first slow-roll parameter and $M=O(10^2 H)$ a high-energy mass scale at which trans-Planckian effects become important \cite{Martin:2003kp,Greene:2004fln}. The parameter space ranges from high frequencies $\om=O(\e^{-1})=O(10^2)$ \cite{Martin:2000xs} or $\om=M/H=O(10^2)$ \cite{Jackson:2013mka} to frequencies $\om=\epsilon/(H/M)=O(1)$ \cite{Martin:2003kp,Greene:2004fln} but very low amplitudes $A^{\rm c}_1=(H/M)^3\e=O(10^{-8})$ and $A^{\rm s}_1=(H/M)^3=O(10^{-6})$ or $A^{\rm c}_1=(H/M)\e=O(10^{-3})$ and $A^{\rm s}_1=H/M=O(10^{-2})$.

None of these trans-Planckian models can be verified by present and next-generation collaborations dedicated to GW astronomy. In fact, not only do these tensor amplitudes become exceptionally low at the frequency scales of GW interferometers, but also, in general, the primordial tensor spectrum is red tilted, i.e., $n_{\rm t}<0$, and cannot generate a \GWB reaching the sensitivity curve of such instruments.


\subsection{Multi-fractional inflation}\label{mufr}

A concrete early-universe model with a spacetime broken DSI is multi-scale inflation. This single-field model of early-time acceleration assumes that spacetime geometry changes with the probed scale, an effect typically found in theories of quantum gravity as pointed out above. A class of theories called multi-fractional spacetimes encode this feature at a fundamental level as a non-trivial integration measure and modified derivative operators in the action \cite{Calcagni:2016azd}. The inflationary spectra of one of these theories, specifically with so-called $q$-derivatives, have been worked out explicitly \cite{Calcagni:2013yqa,Calcagni:2016ofu,Calcagni:2017via}. In particular, the tensor spectrum ${\cal P}_{\rm t}(k)$ features logarithmic oscillations coming from a DSI of the measure in a certain regime of times and distances. In contrast with the models reviewed in sections \ref{revi} and \ref{revi1}, here DSI emerges from the fact that spacetime itself is endowed with one or more characteristic scales. Then, log oscillations affect the scalar and the tensor spectrum alike.

In this model, the primordial spectrum $\cP_{\rm t}(k)$ was calculated in \cite{Calcagni:2020tvw,Calcagni:2016ofu}. Since the cosmic evolution is standard after inflation, the transfer function $\cT$ is approximately constant at the frequencies of GW interferometers and the spectral shape is simply proportional to $\cP_{\rm t}[f/(2\pi)]$. Therefore,
\ba
\hspace{-.8cm}\Om_{\textsc{gw}}(f) &=& \Om_{{\textsc{gw}}*}\left[\frac{\tilde f(f)}{f_0}\right]^{\tilde n_{\rm t}},\label{Omfin}\\
\hspace{-.8cm}\tilde f(f) &=&
f\left\{1+\frac{1}{|\a|}\left(\frac{f}{f_*}\right)^{1-\a}\left[B_0+\sum_{l=1}^{+\infty} \tilde F_l(f)\right]+\dots\right\}^{-1},\label{eq:finve}
\ea
where $\Om_{{\textsc{gw}}*}$ is a constant (notice that $\Om_{{\textsc{gw}}*}\neq \Om_{\textsc{gw}}(f_0)$), $\tilde n_{\rm t}$ is the tensor index of the same inflationary model corresponding to an ordinary spacetime and $\tilde F_l(f)$ is given by \Eq{master2} with different amplitudes $B^{\rm c,s}_l$. Note that a pattern in time opposite with respect to \Eq{eq:finve} holds, since the variable $\tilde t(t)=1/\tilde f(1/t)$ is conjugate to $\tilde f(f)$. The ellipsis indicates other terms of the same functional form as the one showed but with different values of the parameters $\a$, $f_*$, $B_0$, and so on. Regarding the parameters, $\a$ is real and related to the Hausdorff dimension of spacetime, $B_0=0,1$ and the oscillation frequency $\om=2\pi\a/\ln {\rm N}$ has an upper bound at ${\rm N}=2$, where ${\rm N}$ is the number of copies of the underlying fractal (i.e., the fractal spacetime is given by the union of ${\rm N}$ copies of itself rescaled by a factor $\la_\om$). Although it is not a prediction of the theory, it has been argued that oscillations are usually damped and the amplitudes are parametrized by an exponential and/or a power-law decay \cite{Calcagni:2017via}:
\be
B^{\rm c}_l=a^{\rm c}_l \frac{\rme^{-\g l}}{l^u},\qquad B^{\rm s}_l=a^{\rm s}_l \frac{\rme^{-\g l}}{l^u}\,,
\ee
where $\g>0$ and $u>0$. If the amplitudes decay, then the approximation $l_{\rm max}=1$ is acceptable, otherwise one should take more harmonics into account. A simple \emph{Ansatz} is to consider several harmonics with constant amplitudes, e.g., $l_{\rm max}=10$ and $B^{\rm c}_l=B^{\rm s}_l={\rm const}$.

Ignoring log oscillations, \Eq{Omfin} reduces to a double power law,
\be
\Om_{\textsc{gw}}(f)\simeq \Om_{{\textsc{gw}}*}\left(\frac{f_*}{f_0}\right)^{\tilde n_{\rm t}}\left[\frac{1}{\left(\frac{f}{f_*}\right)^{-1}+\frac{B_0}{|\a|}\left(\frac{f}{f_*}\right)^{-\a}}\right]^{\tilde n_{\rm t}},
\ee
which was studied in \cite{Calcagni:2020tvw}. However, here we want to make a different approximation and retain the log-periodic pattern. At the high frequencies typical of present- and next-generation GW interferometers ($f_*\ll f_0$), the power-law term in \Eq{eq:finve} dominates:
\be
\hspace{-.8cm}\Om_{\textsc{gw}}(f) \simeq \tilde\Om_{\textsc{gw}*} \left(\frac{f}{f_0}\right)^{\a \tilde n_{\rm t}}\left[B_0+\sum_{n=1}^{+\infty} \tilde F_n(f)\right]^{-\tilde n_{\rm t}},
\ee
where $\tilde\Om_{\textsc{gw}*}=|\a|\Om_{\textsc{gw}*}(f_0/f_*)^{\a-1}$ and $\tilde n_{\rm t}$ is small and negative as in standard inflation. The power of log oscillations can be approximated to a linear dependence on the oscillations themselves, since these are bounded from above and from below and the amplitudes $B^{\rm c}_l$ and $B^{\rm s}_l$ cannot be too large lest they generate excessively sharply peaked features. If $|B^{\rm c,s}_l|\lesssim 1$ for all $l$, then for $B_0=1$ 
\ba
\hspace{-1.5cm}&&\left[f^\a\left(1+\sum_l B^{\rm c}_l\cos\ln f^{l\om} + \sum_lB^{\rm s}_l\sin \ln f^{l\om}\right)^{-1}\right]^{\tilde n_{\rm t}}\nonumber\\
\hspace{-1.5cm}&&\qquad\qquad\simeq f^{\a \tilde n_{\rm t}}(1+A^{\rm c}_1\,\cos\ln f^{\omega} + A^{\rm s}_1\,\sin \ln f^{\omega}+\dots)\,,\\
\hspace{-1.5cm}&& A^{\rm c}_1=-\tilde n_{\rm t}B^{\rm c}_1\,,\quad A^{\rm s}_1=-\tilde n_{\rm t}B^{\rm s}_1\,,
\ea 
thus reproducing \Eq{master2}. Since the theoretical spectral index is small, $\tilde n_{\rm t}\ll 1$, so are the amplitudes $A^{\rm c}_1$ and $A^{\rm s}_2$. 

In cosmological applications to date, values of $\a<0$ have been considered in order to get a blue-tilted spectrum at frequencies higher than the CMB range \cite{Calcagni:2020tvw}. In that case, the predicted amplitude of the \GWB intersects the DECIGO sensitivity curve but does not reach the sensitivity curves of LISA and ET for $|\a|=O(1)$ \cite{Calcagni:2020tvw}. We checked that the \GWB of this model can be lifted up to the LISA and ET ranges provided $\a\lesssim-30$. These values of $\a$ are not welcome theoretically because the Hausdorff dimension of spacetime (i.e., the scaling of 4-volumes with their linear size) is $d_\textsc{h}=\alpha_0+3\a$ and it must be positive definite. Large negative values of $\a$ would imply that the Hausdorff dimension of time $\a_0$ be very large at the frequency scale $f_*$ where the $\a$ power dominates, which is not a natural geometric configuration.

For phenomenological purposes, one could get along with this possibility and explore its consequences. However, it is difficult to support inflation with strongly negative values of $\a$, in which case the observed tensor index $n_{\rm t}=\a \tilde n_{\rm t}$ would be positive and large. The reason is that the observed scalar spectral index $n_{\rm s}-1=(\tilde n_{\rm s}-1)\a$ \cite{Calcagni:2016ofu,Calcagni:2020tvw} can be negative and small only if the scalar index $\tilde n_{\rm s}-1$ of the corresponding inflationary model in Einstein gravity is positive and very small. Although there may be models with these characteristics (e.g., multi-field inflation), they could entail a moderate level of fine tuning. 

Therefore, it is more natural to consider the case where the logarithmically modulated power-law term in \Eq{eq:finve} relevant at the frequencies of GW interferometers is not the same one important at CMB scales, which would then be relegated to the ellipsis. As we said, the latter stands for other powers, modulated by log oscillations, with different scales $f_*\to f_{**},\ldots$, exponents $\a\to\a_{**},\ldots$, frequency parameters $\om\to\om_{**},\ldots$ and amplitudes $A_l^{\rm c,s}\to A_l^{{\rm c,s}**},\ldots$. These extra powers $(f/f_{**})^{1-\a_{**}}$, $\ldots$, exist in multi-fractal geometries with three or more inequivalent regimes, separated by two or more length (energy, frequency) scales. In this case, assuming the \GWB to be steep enough in its rise in the intermediate regime with a negative $\a_{**}<-30$, then $\a$ in \Eq{eq:finve} can even be positive and the spectral shape still cross the LISA or ET sensitivity curve. Then, none of the constraints coming from the CMB \cite{Calcagni:2016ofu,Braglia:2021rej} apply to the parameters $\a$ and $f_*$ or to the amplitudes $A_l^{\rm c,s}$ of the log oscillations. At any rate, we will see that log-periodic features are detectable also for amplitudes smaller than the CMB upper bound $A_l^{\rm c,s}< 0.4$ \cite{Calcagni:2016ofu}.

This conclusion holds for any cosmological model coming from quantum gravity with an intrinsic DSI: If there are at least three characteristic spacetime scales such that the intermediate one is associated with a negative anomalous scaling, then one can avoid CMB constraints on log oscillations and use the spectral shape \Eq{master1}--\Eq{master2} just at GW interferometer frequencies.


\section{Prospect for detection in next-generation GW interferometers}\label{inter}

Having motivated the spectral shape given by equations (\ref{master1})--(\ref{master2}) on general grounds in section \ref{maindsi} and with several general mechanisms illustrated by the examples in sections \ref{revi} and \ref{dsiqg}, we now explore the detectability of log oscillations using next-generation GW interferometers. The search for a \GWB is performed by cross-correlating signals from multiple interferometers. For a given set of noise spectra $S_{I,J}(f)$ from different interferometers $I$ and $J$, the expected signal-to-noise ratio (SNR) is given by~\cite{Allen:1997ad}
\begin{equation}
\text{SNR}^2 = \left(\frac{3H_0^2}{10\pi^2}\right)^2 2T_{\text{obs}}\sum_{I,J}\int_0^{+\infty}\rmd f\frac{|\gamma_{IJ}(f)|^2\Om_\textsc{gw}(f)^2}{f^6S_I(f)S_J(f)},
\label{SNR}
\end{equation}
where $T_{\text{obs}}$ is the observation time, $\gamma_{IJ}$ is the overlap reduction function, $H_0=100h~\text{km}~ \text{s}^{-1}\text{Mpc}^{-1}$ and we adopt the value from the \textsc{Planck} satellite, $h=0.67$~\cite{Planck:2018jri}. The overlap reduction function is given by integrating the contributions from all directions, obtained using the detector responses $F^+_{I,J}$ and $F^\times_{I,J}$:
\begin{equation}
\gamma_{IJ}(f) \coloneqq \frac{5}{8\pi}\int \rmd\hat{\bf \Omega}\, (F^+_IF^+_J+F^{\times}_IF^{\times}_J)\,\rme^{-2\pi \rmi f\hat{\bf \Omega}\cdot ({\bm x}_I-{\bm x}_J)},
\end{equation}
where $\hat{\bf \Omega}$ is the solid angle and ${\bm x}_{I,J}$ denote the positions of the detectors.

The Fisher matrix ${\cal F}_{ij}$ is a widely used tool to forecast the performance of future experiments. Under the assumption that the likelihood function can be well approximated by a Gaussian distribution near the maximum likelihood estimate of the parameters, the inverse of the Fisher matrix provides a lower bound on the expected error $\sigma_{p_i}=\sqrt{(\cF^{-1})_{ii}}$. For the cross-correlation analysis of a GW data set, the Fisher matrix is given by~\cite{Seto:2005qy,Kuroyanagi:2009br}
\begin{equation}
{\cal F}_{ij} = \left(\frac{3H_0^2}{10\pi^2}\right)^2 2T_{\text{obs}}\sum_{I,J}\int_0^{+\infty}\rmd f\frac{|\gamma_{IJ}(f)|^2\partial_{p_i}\Omega_{\textsc{gw}}(f)\partial_{p_j}\Omega_{\textsc{gw}}(f)}{f^6S_I(f)S_J(f)}.
\label{Fisher}
\end{equation}
This formula may be slightly different for different observatories such as ET (whose design is still in consultation~\cite{Branchesi:2023mws}) and LISA, where the auto-correlation method is planned for the search of a \GWBn. However, the actual formula is similar and all the arguments in the following subsections hold for all these cases.

\begin{figure}[!ht]
    \centering
    \includegraphics[keepaspectratio, scale=0.8]{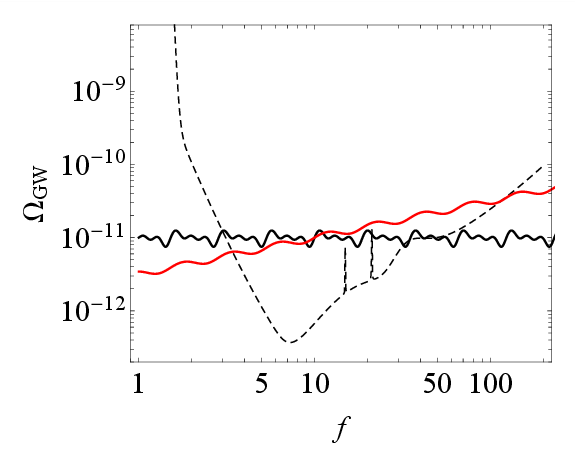} 
    \caption{Examples of a \GWB spectrum with logarithmic oscillations, displayed with the sensitivity curve of ET. The red curve depicts the $l_{\rm max}=1$ case, plotted using equation \eqref{monotonic} with $\Omega_0=10^{-11}$, $n_{\rm t}=0.5$, $\omega=10$, $A_1=0.1$, $\Phi_1=0$. The black curve represents the $l_{\rm max}=3$ case, plotted using equation \eqref{higher} with $\Omega_0=10^{-11}$, $n_{\rm t}=0$, $\omega=10$, $A_1=0.1$, $\Phi_1=0$, $A_2=0.1$, $\Phi_2=0$, $A_3=0.1$, and $\Phi_3=0$. The pivot scale is set at $f_*=10$ Hz.}
    \label{Fig1}
\end{figure}

Using this formalism, we can estimate the expected error for a given sensitivity curve. In the following, to provide an example, we discuss the constraining power using the sensitivity of the ET. In figure \ref{Fig1}, we show examples of the \GWB signal (for $l_{\rm max}=1$ and $3$) and the ET sensitivity, plotted with the criterion of having SNR $=1$ for each bin $\Delta\ln f = 1/10$. We assume a $3$-year observation with the sensitivity given by the ET-D curve taken from~\cite{ETsensitivity}. For simplicity, we assume that the overlap reduction function is $|\gamma_{IJ}|=1$, which is the case for co-located and co-aligned L-shaped detectors. The result simply rescales by $|\gamma_{IJ}|=3/8$ for the triangular configuration, where the detectors form a V-shape with arm angles of 60 degrees and a separation of 120 degrees \cite{Regimbau:2012ir}. In the context of ET, it is a reasonable approximation to consider $|\gamma_{IJ}(f)|={\rm const}$ for frequencies of interest, $f<10^3$~Hz.


\subsection{One harmonic}

We start with the minimal $l_{\rm max}=1$ case assuming a power-law form of the basic \GWB shape. In this case, the generic form of the \GWB spectral amplitude \Eq{master1} can be written as
\be
\Omega_{\textsc{gw}}(f)=\Omega_0\left(\frac{f}{f_*}\right)^{n_{\rm t}}\left[1+A_1\sin\left(\omega\ln{\frac{f}{f_*}}+\Phi_1\right)\right].
\label{monotonic}
\ee
Note that we use the oscillation amplitude $A_1$ and the phase $\Phi_1$, rather than two separate oscillation amplitudes $A_1^{\rm c}$ and $A_1^{\rm s}$ for cosine and sine. This parametrization may have advantages in computation time when performing searches in the parameter space. Specifically, the prior range of the phase can be limited to $-\pi \leq \Phi_1 \leq \pi$, while we typically want to explore a large parameter space for the oscillation amplitude ranging in log scale. Having a flat prior on $\Phi_1$ with the parametrization \eqref{monotonic} means that we assign less importance to fine-tuned modulations such as in the cases where $A_1^{\rm c} \ll A_1^{\rm s}$ or $A_1^{\rm c} \gg A_1^{\rm s}$. 

In the $l_{\rm max}=1$ case, we have 5 free parameters to determine in parameter estimation, namely, $p_i=[\Omega_0, n_{\rm t}, \omega, A_1, \Phi_1]$. The pivot frequency $f_*$ can be taken to be arbitrary without loss of generality (a change in $f_*$ can be absorbed by changes in the normalization amplitude $\Omega_0$ and the oscillation phase $\Phi_1$). Here we take $f_*=10$~Hz to explore the case of ET, as it is convenient to set it around the frequency band of the observatory.

\begin{figure}[!ht]
    \centering
    \includegraphics[keepaspectratio, scale=0.65]{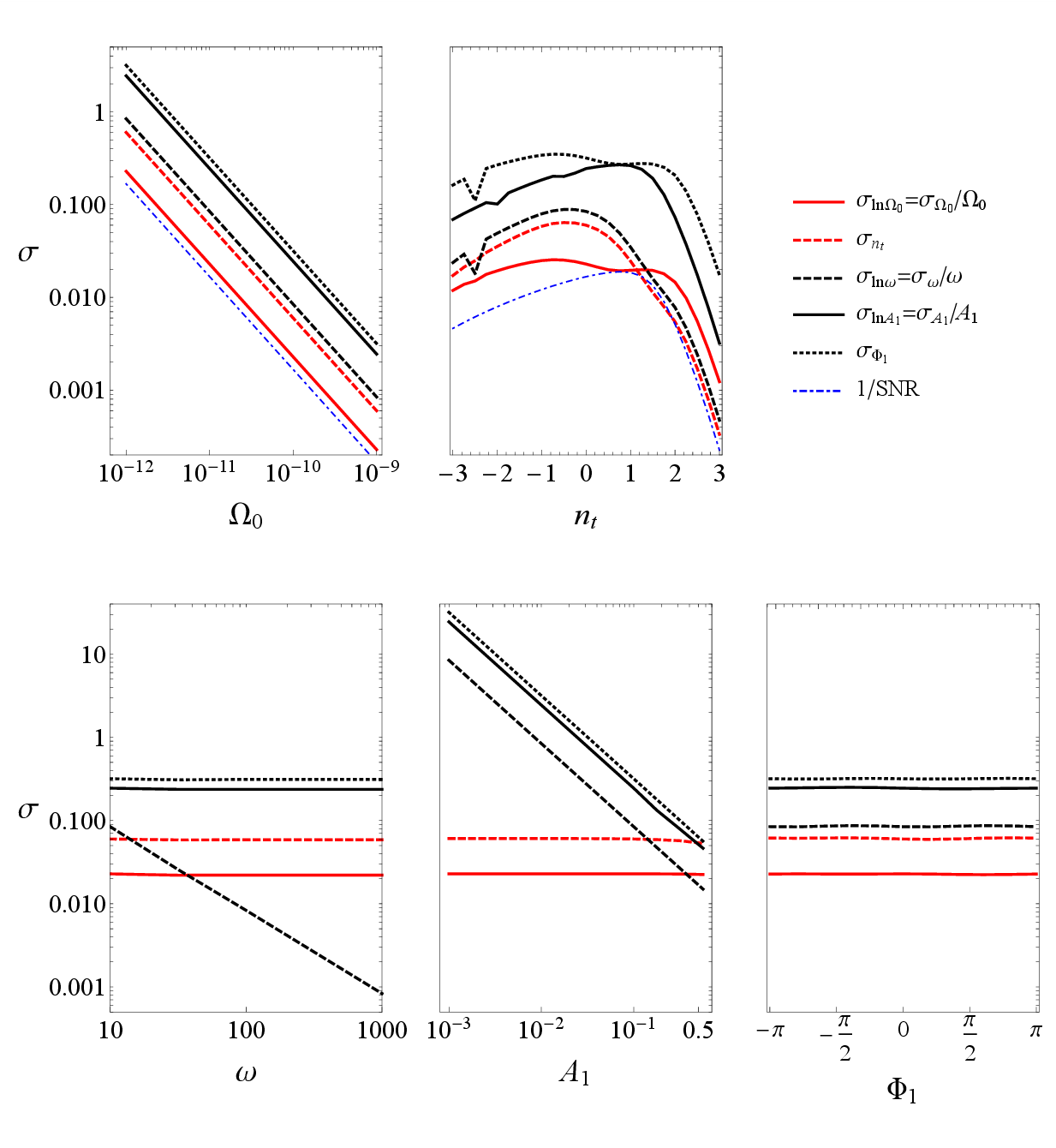} 
    \caption{Parameter dependence of the expected errors $\sigma$. The red curves describe errors on the parameters that determine the envelope of the spectrum, such as the normalization amplitude $\Omega_0$ (red solid) and the tilt $n_{\rm t}$ (red dashed). The black curves correspond to the errors on oscillation parameters, such as the oscillation frequency $\omega$ (black dashed), the oscillation amplitude $A_1$ (black solid) and the oscillation phase $\Phi_1$ (black dotted). Each figure shows the errors by varying one parameter, while the other parameter values are fixed at $\Omega_0=10^{-11}$, $n_{\rm t}=0$, $\omega=10$, $A_1=0.1$ and $\Phi_1=0$. For reference, we also show the $1/\text{SNR}$ curve (blue dot-dashed).}
    \label{Fig2}
\end{figure}

In figure \ref{Fig2}, we present the parameter dependence of the expected errors assuming the ET sensitivity. For parameters that could vary over several orders of magnitude, we express the errors as relative errors, i.e., errors on the logarithm of the value, denoted as $\sigma_{\ln p_i}$. This allows us to roughly set a common threshold of $\sigma < 1$ for all parameters to judge whether they can be constrained by data. When we require an accurate determination of the parameters, we may set a lower threshold $\sigma < 0.1$.

The top panels depict the dependence on parameters that determine the envelope of the spectrum, namely the normalization amplitude $\Omega_0$ and the tilt $n_{\rm t}$. We observe a trend where the errors for all parameters scale as $\propto 1/\Omega_0$. This is reasonable, since parameter determination becomes more straightforward for a \GWB signal with higher SNR. The dependence on $n_{\rm t}$ exhibits a non-trivial shape, reflecting the sensitivity curve of the observatory and roughly following a trend $\propto 1/\text{SNR}$, as shown by the inclusion of the $1/\text{SNR}$ curve for reference.

The bottom panels illustrate the dependence on oscillation parameters. We observe that the error on $\ln\omega$ decreases for high-frequency oscillations, and the errors on all oscillation parameters improve with larger oscillation amplitude. This can be interpreted as the fact that having more oscillations within the observation frequency band as well as a larger oscillation amplitude help to determine the oscillation parameters more precisely. We present an analytic estimation for understanding these parameter dependencies in \ref{appB}. Additionally, we find that the determination of the parameters $\Omega_0$ and $n_{\rm t}$ for the envelope remains unaffected by changes in the oscillation parameters. However, note that this is not the case when $\om < 10$, since for small $\om$ frequencies the oscillation period becomes comparable with, or larger than, the ET frequency sensitivity band. In such cases, we find a strong degeneracy among the parameters and the Fisher method fails to provide accurate estimates.

While we presented this example assuming the sensitivity of ET, the above trends should hold for other observatories as well. Generally speaking, $\sigma\propto 1/\text{SNR}$ so the higher the signal-to-noise ratio, the smaller the magnitude of the errors for all parameters. The SNR is mainly determined by the sensitivity curve of the observatory and the envelope of the \GWB spectrum, controlled by $\Omega_0$ and $n_{\rm t}$. Thus, the general statement we can make for all interferometer-type observatories is that log oscillations can be detected if the \GWB detection is made with SNR $\gtrsim 100$ for $\omega \sim 10$ and $A_1\sim 0.1$. Another generic trend is that the errors on oscillation parameters scale as $\sigma_{\ln \omega}\propto 1/(A_1\omega)$, $\sigma_{\ln A_1}\propto 1/A_1$ and $\sigma_{\Phi_1}\propto 1/A_1$, as we confirm analytically in \ref{appB}. Furthermore, even though we parametrized the \GWB envelope using the power-law form, this argument can be extended to any envelope modulated by small oscillations.

The only difference that could arise from changing observatory is the accessible range of $\omega$. There exists an experimental upper bound on $\omega$ due to the frequency resolution of the instrument. Given a frequency resolution $\Delta \ln f$ in log space, the observatory can resolve the log-oscillatory signal when
\begin{equation}
\label{eq:upbo}
\om< \frac{1}{\Delta \ln f} = \frac{f}{\Delta f}\,.
\end{equation}
The frequency resolution is given by $\Delta f = 1/T_{\rm obs} = f_{\rm s}/N$, where $f_{\rm s}$ is the sampling frequency of the experiment, and $N$ is the number of data points. For example, in the LVK analysis~\cite{KAGRA:2021kbb}, the sampling frequency is $16384$~Hz and the down-sampled data of $f_{\rm s}=4096$~Hz is used for the stochastic analysis. One year of data cannot be realistically analyzed as a whole because of the computation power needed to process data points of order of $N=f_{\rm s} T_{\rm obs} \sim 10^{11}$, as well as because of disruptions of the data flow due, e.g., to glitch noise or run breaks. Therefore, the data are usually split into segments of $192$ seconds, which gives a frequency resolution of $\Delta f \approx 0.0052$~Hz further coarse-grained to $\Delta f = 1/32\approx 0.031$~Hz. This yields a resolution of $\Delta \ln f \approx 1/320$ around $10$~Hz for the coarse-grained data. A higher resolution could be achieved by taking the segment size longer without applying coarse-graining. In principle, $\Delta \ln f = O(10^{-9})$ around $10$~Hz can be achieved if we can analyze one year of data at once. A similar argument would hold for ET and we would have $\Delta \ln f = O(10^{-3})$ unless the computation power and data analysis techniques were dramatically improved by the time of the observation run. In summary, the range of observable $\omega$ values is constrained by factors such as the data's maximum duration after being affected by noise and run interruptions, as well as the computational time required for processing numerous data points. Nevertheless, a frequency resolution of $\Delta \ln f = O(10^{-3})$ is good enough compared to the typical width of the oscillations in log space.

In the context of LISA, the mission aims to explore a lower frequency band around $\sim 10^{-3}$ Hz. Due to periodic antenna re-pointing and operational interruptions, the data will likely be divided into segments of approximately 10 days each~\cite{Flauger:2020qyi}. This means that, without coarse-graining, achieving a resolution of $\Delta \ln f = O(10^{-4})$ around $10^{-3}$ Hz should be feasible. LISA is designed with a much lower sampling frequency of $f_{\rm s} = 3.3$ Hz, resulting in a total data point count of approximately $N \approx 3 \times 10^{8}$. Consequently, this set-up should be more manageable in terms of the required computational power.

On the other hand, there is also a lower bound on $\om$. As mentioned above, the oscillation effect starts to degenerate with the entire spectral shape when $\omega$ is too small. To avoid the degeneracy, we need at least one oscillation cycle within the sensitivity band. This requirement strongly depends on the shape of the sensitivity curve and on the amplitude of the \GWBn, as the observable bandwidth tends to widen when the \GWB signal has a high amplitude. Roughly speaking, we need $\omega > 10$ to have enough oscillation cycles inside the sensitivity band. This condition would be relaxed when $\Omega_0$ is higher.
\begin{table}[t]
  \centering
  \begin{tabular}{llccc}
    \hline
    Experiment  & Peak sensitivity frequency & $\Omega_0$  & $n_{\rm t}$ &  $\omega$\\
    \hline \hline
    ET/CE  & $1\!-\!10$\,Hz & $\gtrsim 10^{-11}$  & $\gtrsim 0.28$ & $10 < \omega < 10^3$ ($10^9$) \\
    DECIGO & $0.1\!-\!1$\,Hz  &  $\gtrsim 10^{-15}$  & $\gtrsim 0.06$ & $10 < \omega < ~~?~~ $ ($10^7$) \\
    LISA  & $10^{-3}\!-\!10^{-2}$\,Hz  &  $\gtrsim 10^{-11}$  & $\gtrsim 0.34$ & $10 < \omega < 10^4$ ($10^5$) \\ 
    \hline
  \end{tabular}
  \caption{The table shows the minimum \GWB amplitude $\Omega_0$ required to detect log oscillations with an amplitude $A_1\sim 0.1$ as well as the value of the minimal tilt $n_{\rm t}$ needed to achieve such amplitude for a \GWB having the current upper-bound value $r=0.036$ of the tensor-to-scalar ratio at the CMB scale $k=0.05\,{\rm Mpc}^{-1}$. The value of the normalization amplitude $\Omega_0$ is estimated assuming that SNR $\gtrsim 100$ is needed for a precise determination of the oscillation parameters with $\sigma \sim 0.1$. Additionally, we provide the rough range of $\omega$ accessible by the instrument. In brackets, we put the ideal upper bound achieved if one analyzed the data taking the entire observation time as a whole (see the main text for details).\label{table:experiment}}
\end{table}

In table~\ref{table:experiment}, we summarize the prospect for future GW observatories. The table contains the target frequency of each observatory, the required amplitude for the determination of the oscillation parameters when $A_1\sim 0.1$, the value of the minimal tilt $n_{\rm t}$ needed to achieve such amplitude for a \GWB having the current upper-bound value $r=0.036$ of the tensor-to-scalar ratio at the CMB scale $k=0.05\,{\rm Mpc}^{-1}$~\cite{BICEP:2021xfz}, and the explorable range of $\omega$.

Note that, from what said at the end of section \ref{mufr}, if we assume the presence of an intermediate stage between CMB and interferometer scales, we do not have to take into account CMB constraints on the oscillation amplitudes (which, at least for the case of multi-fractional inflation, does not reduce the parameter ranges at high frequencies anyway).


\subsection{Multiple harmonics}

Now we evaluate the case where we include higher harmonics and the \GWB spectral amplitude \Eq{master1} gets more terms as
\ba
\fl\Omega_{\textsc{gw}}(f)&=&\Omega_0\left(\frac{f}{f_*}\right)^{n_{\rm t}}\left[1
+A_1\sin\left(\omega\ln{\frac{f}{f_*}}+\Phi_1\right)\right. \nonumber\\
\fl&&\hspace{2.1cm}\left.+A_2\sin\left(2\omega\ln{\frac{f}{f_*}}+\Phi_2\right)+A_3\sin\left(3\omega\ln{\frac{f}{f_*}}+\Phi_3\right)+\cdots\right].
\label{higher}
\ea
As seen from the equation, each time one adds a higher-order term, the number of free parameters increases by two (oscillation amplitude $A_l$ and phase $\Phi_l$). 

\begin{figure}[!ht]
    \centering
    \includegraphics[keepaspectratio, scale=0.6]{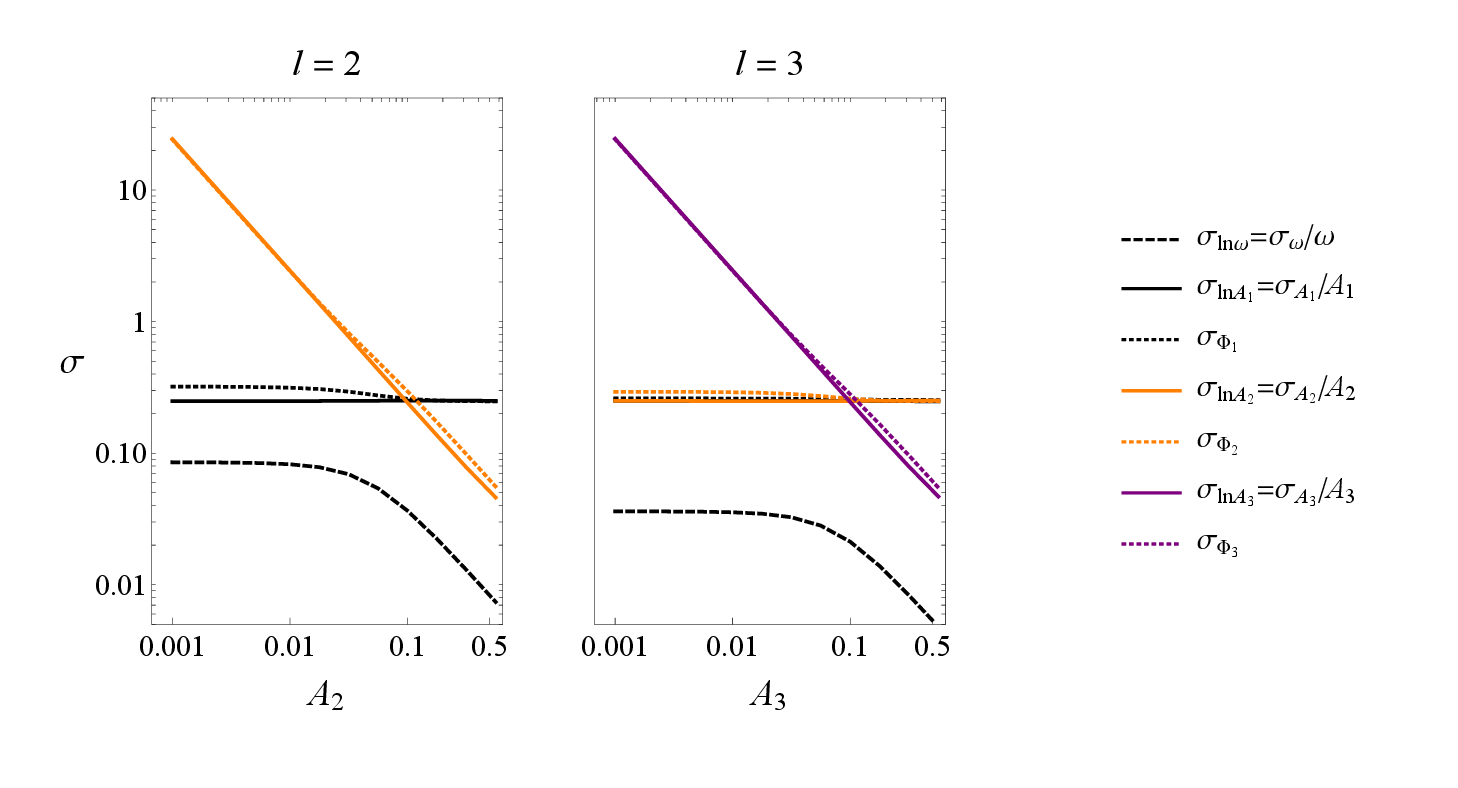} 
    \caption{Dependence of the expected errors $\sigma$ for multiple harmonics, where we include up to $l_{\rm max}=2$ (left) and $l_{\rm max}=3$ (right), plotted by varying the oscillation amplitude $A_2$ and $A_3$, respectively. The other parameters are fixed at $\Omega_0=10^{-11}$, $n_{\rm t}=0$, $\omega=10$, $A_1=0.1$, $\Phi_1=0$ for the $l_{\rm max}=2$ case, while for the $l_{\rm max}=3$ case the additional parameters are $A_2=0.1$, $\Phi_2=0$ and $\Phi_3=0$. The black, orange and purple curves correspond to the oscillation parameters (oscillation amplitude $A_l$ and phase $\Phi_l$) for $l=1$, $l=2$, and $l=3$, respectively.}
    \label{Fig3}
\end{figure}
In figure \ref{Fig3}, we display the expected errors for the cases where we include up to $l_{\rm max}=2$ (left) and $l_{\rm max}=3$ (right). Each panel is plotted by varying the oscillation amplitude of the higher-order mode, $A_2$ and $A_3$, respectively. Surprisingly, we find that adding higher-order terms does not affect the estimation of lower-order terms. For instance, when comparing the left and right panels, we observe that the error magnitude of the $l=1$ parameters ($\sigma_{\ln A_1}$ and $\sigma_{\Phi_1}$) remains the same, even after including the $l=3$ mode (notably, the orange solid and black solid curves completely overlap). Moreover, we can see that the errors are largely insensitive to the magnitude of the additional harmonics. Interestingly, the errors on $\omega$ actually improve when more harmonics are added and they decrease further when higher-order oscillations have larger amplitudes (i.e., when $A_2 > A_1$ and $A_3 > A_2$). Although it is not shown in the figure in order to avoid including too many lines, we have also confirmed that the addition of higher-order modes does not affect the estimations of $\Omega_0$ and $n_{\rm t}$.

This behavior may seem contrary to one's expectations, as errors typically increase when more free parameters are added. The reason behind this is that the oscillation frequency of each term shares the same parameter $\omega$ and each term contributes independent information with oscillations of $\omega$, $2\omega$, $3\omega$ and so on. As each harmonic is independent, the error on each mode remains unaffected even after the marginalization process. Moreover, by including higher-order terms, we obtain more information about $\omega$, leading to a reduction in the error. Additionally, the error decreases for larger oscillation amplitudes of higher harmonics because having more oscillations within the sensitivity band helps to improve accuracy. 

We stop the investigation at $l_{\rm max}=3$ but this tendency remains consistent when adding more harmonics. This implies that including additional harmonics is advantageous for obtaining more information. However, one should also consider that adding more parameters increases the computation time required for the parameter search. Therefore, the number of higher-order terms to be included in the analysis should be carefully chosen depending on the model one intends to explore and its theoretical prediction for higher-order modes.


\section{Conclusions}\label{concl}

In this paper, we have shown that the spectral shape \Eq{master1}--\Eq{master2}, given by a multi-harmonic linear superposition of log-periodic functions, is the most general spectral shape of a \GWB displaying log oscillations in a given range of GW frequencies $f$, independently of its physical origin. This can be a valuable tool to control analytically log-oscillating models for which the \GWB has been computed only numerically or in open form. We have reviewed log-oscillating primordial spectra and stochastic backgrounds originated either by second-order scalar perturbations or by tensor modes and also discussed new ways to obtain a log-periodic \GWBn, in particular, in any model of (or motivated by) quantum gravity endowed with a discrete scale symmetry. In these cases, DSI originates from modified dispersion relations or Fock vacua or, at a more fundamental level, as the manifestation of self-similar spacetime geometries.

We also investigated the detectability of log-oscillation features in future GW observatories by calculating the Fisher matrix in the case of the Einstein Telescope. The main result we obtained from the analysis is that we can determine the oscillation parameters with an $O(10\%)$ precision if we have a detection with SNR $\gtrsim 100$ when the oscillation amplitudes are $A_l\sim 0.1$ and the frequency is $\om\sim 10$. The errors generically decrease as $\propto 1/{\rm SNR}$ for all parameters. Also, the errors on oscillation parameters scale as $\sigma_{\ln \omega}\propto 1/(A_l\omega)$, $\sigma_{\ln A_l}\propto 1/A_l$ and $\sigma_{\Phi_l}\propto 1/A_l$. This result generically applies also to other missions such as CE, LISA and DECIGO; the prospect for different types of observation is summarized in table~\ref{table:experiment}. Another relevant finding is that the error is insensitive to the inclusion of higher-order harmonics, since each term can be constrained independently as the oscillation frequencies are ordered hierarchically as integer multiples of $\om$.

An interesting theoretical aspect to explore in the future is the possibility that DSI is actually a gauge symmetry, thus respecting, in its exact form, the conjecture that quantum gravity does not possess global symmetries \cite{Banks:2010zn,Harlow:2018jwu,Harlow:2018tng}. Even if, to date, there is no compelling rigorous argument supporting such conjecture in theories of quantum gravity not embedded in string theory or the AdS/CFT correspondence, one could temporarily assume its validity and see how one can gauge DSI and what the physical consequences of such operation are.

Logarithmic oscillations are just one more feature of models of the early universe within and beyond Einstein gravity that can be systematically explored with gravitational waves. In the latter case, GW observations are a promising window into the intimate nature of the gravitational interaction.

Finally, we also mention that our results can be easily extended beyond signals with logarithmic oscillations. The proof in \ref{appA} holds also under the replacement $\ln(f/f_*)\to f/f_*$, so that we can conclude that the most general spectral shape with \emph{linear} oscillations is
\be
\Om_{\textsc{gw}}(f) = \bar\Om_{\textsc{gw}}(f)\left\{1+\sum_{l=1}^{l_{\rm max}} \left[A^{\rm c}_l\cos\left(l\om\frac{f}{f_{*}}\right)+A^{\rm s}_l\sin\left(l\om\frac{f}{f_{*}}\right)\right]\right\}.\label{masterlin}
\ee
This expression could be used to fit the signal predicted in several inflationary models (for instance, with sharp features) \cite{Achucarro:2010da,Adams:2001vc,Bean:2008na,Park:2012rh,Miranda:2012rm,Bartolo:2013exa,Hazra:2014jka,Hazra:2014goa,Palma:2014hra,Hazra:2016fkm,Ballesteros:2018wlw,Palma:2020ejf,Kefala:2020xsx,Fumagalli:2020nvq,Dalianis:2021iig,Boutivas:2022qtl} where, most of the times, the power spectra or the stochastic background are only depicted in figures without providing a closed-form expression such as \Eq{masterlin}.


\section*{Acknowledgments}

GC is supported by grant PID2020-118159GB-C41 funded by MCIN/AEI/10.13039/ 501100011033 and thanks Johannes Th\"urigen and Xingang Chen for useful e-mail discussions. SK is supported by the Atracci\'on de Talento contract 2019-T1/TIC-13177 from Comunidad de Madrid, the I+D grant PID2020-118159GA-C42 and the Spanish Research Agency (Agencia Estatal de Investigaci\'on) through the Grant IFT Centro de Excelencia Severo Ochoa CEX2020-001007-S, both funded by MCIN/AEI/10.13039/501100011033, the i-LINK 2021 grant LINKA20416 of CSIC, and Japan Society for the Promotion of Science (JSPS) KAKENHI Grants 20H01899, 20H05853 and 23H00110.


\appendix


\section{Universality of the log-oscillating spectral shape}\label{appA}

In this appendix, we prove that, if a \GWB displays logarithmic oscillations in some range of frequencies $f$, then it must follow the spectral shape \Eq{master1}--\Eq{master2} in that range. To see this, we only use symmetry and several trigonometric identities. We make only one assumption to make the proof work.

Consider the generic spectral shape
\be\label{omgen}
\Om_{\textsc{gw}}(f) = \bar\Om_{\textsc{gw}}(f)\,\Psi(f)\,,
\ee
where $\bar\Om_{\textsc{gw}}(f)$ depends on the model and $\Psi$ is an arbitrary real-valued function of the GW frequency. Set $f_*=1$ to simplify the following expressions. By definition, the presence of log oscillations of a certain frequency $\om$ (not to be confused with the GW frequency $f$) implies a DSI, governed by the transformation law \Eq{eq:dsi}. This DSI must appear in the modulation factor $\Psi$, which then must depend on $f$ via the simplest DSInvariant functions $\Lambda_i(f)$ of log-periodicity $2\pi$:
\be\label{eq:dsiappA}
\Psi(f)=\Psi[\{\Lambda_i(f)\}]\,,\qquad \Lambda_i(\la_\om f)=\Lambda_i(f)\,,\qquad\la_\om=\rme^{\frac{2\pi}{\om}}\,.
\ee
Such functions $\Lambda_i$ are the elementary trigonometric functions of the logarithm:
\be\label{cFfor}
\Psi(f)
=\Psi[\cos(\om\ln f),\sin(\om\ln f)]\,,
\ee
so that $\Psi$ is indeed DSInvariant:
\ba
\Psi(\la_\om f)&=&\Psi\{\cos[\om\ln (\la_\om f)],\sin[\om\ln (\la_\om f)]\}\nn
&=&\Psi[\cos(\om\ln f+2\pi),\sin(\om\ln f+2\pi)]\nn
&=&\Psi[\cos(\om\ln f),\sin(\om\ln f)]\nn
&=&\Psi(f)\,.
\ea
Note that a generalization to the presence of constant phases inside the trigonometric functions leads to the same formal expression \Eq{cFfor}, since a phase only produces a linear combination of sine and cosine, e.g., $\cos(\om\ln f + \Phi) =\cos\Phi\cos (\om\ln f)-\sin\Phi\sin (\om\ln f)$. 

Assuming that $\Psi$ is analytic in its two arguments,\footnote{Therefore, this mini theorem does not encompass non-analytic cases such as $\Psi(f)=1/\sqrt{|\cos(\om\ln f)|}$.} one can Taylor expand it as
\be\label{cF1}
\Psi(f)=\sum_{p=0}^{+\infty}\sum_{q=0}^{+\infty}a_{p,q}\cos^p(\om\ln f)\,\sin^q(\om\ln f)\,,
\ee
where $a_{p,q}$ are some coefficients. An integer power of the sine or the cosine is given by a linear superposition of sines and cosines with their argument multiplied by an integer. Depending on whether $p$ and $q$ are odd or even \cite[formul\ae\ 1.320]{GR},
\ba
\fl\cos^p\theta &=& \left\{\begin{matrix}\hspace{1.4cm} \frac{1}{2^{2n-2}} \sum_{r=0}^{n-1} \binom{2n-1}{r} \cos[(2n-2r-1)\theta]\,, &\hspace{.5cm} p=2n-1\\\hspace{.9cm}
\\
\frac{1}{2^{2n}} \binom{2n}{n} + \frac{1}{2^{2n-1}} \sum_{r=0}^{n-1} \binom{2n}{r} \cos[(2n-2r)\theta]\,, & p=2n\end{matrix}\right.,
\ea
\ba
\fl\sin^q\theta &=& \left\{\begin{matrix} \hspace{.2cm}\frac{1}{2^{2n-2}} \sum_{s=0}^{n-1} (-1)^{n-s-1} \binom{2n-1}{s} \sin[(2n-2s-1)\theta]\,, & \hspace{.5cm} q=2n-1\\
\\
\frac{1}{2^{2n}} \binom{2n}{n} + \frac{1}{2^{2n-1}} \sum_{s=0}^{n-1} (-1)^{n-s} \binom{2n}{s} \cos[(2n-2s)\theta]\,, & q=2n\end{matrix}\right.,
\ea
which allows us to recast \Eq{cF1} as
\ba
\fl\Psi(f) &=& \sum_{n=0}^{+\infty}\sum_{m=0}^{+\infty}\sum_{r=0}^{n-1}\sum_{s=0}^{n-1} \left[b_{n,m,r,s} \cos(t_{n,m,r,s}^{(1)}\om\ln f)\, \cos (t_{n,m,r,s}^{(2)}\om\ln f)\right.\nn
\fl&& \qquad\qquad\qquad\left.+\tilde b_{n,m,r,s} \cos(t_{n,m,r,s}^{(3)}\om\ln f)\,\sin(t_{n,m,r,s}^{(4)}\om\ln f)\right],\label{cF2}
\ea
where $b_{n,m,r,s}$ and $\tilde b_{n,m,r,s}$ are coefficients that can be read from the above formul\ae\ and $t_{n,m,r,s}^{(i)}\in\mathbb{N}$ are linear combinations of $n,m,r,s$ yielding positive integers. But now
\ba
2\cos(t^{(1)}\theta) \cos (t^{(2)}\theta)&=&\cos[(t^{(1)}+t^{(2)})\theta]+\cos[(t^{(1)}-t^{(2)})\theta]\,,\\
2\cos(t^{(3)}\theta) \sin (t^{(4)}\theta)&=&\sin[(t^{(4)}+t^{(3)})\theta]+\sin[(t^{(4)}-t^{(3)})\theta]\,,
\ea
so that, since $t^{(1)}-t^{(2)}$ and $t^{(4)}-t^{(3)}$ can also take negative integer values, we can rewrite \Eq{cF2} as a Laurent series:
\be\label{laur}
\Psi(f)=\sum_{l=-\infty}^{+\infty}\left[\tilde A^{\rm c}_l\cos(l\om\ln f)+\tilde A^{\rm s}_l\sin(l\om\ln f)\right].
\ee
This reduction to just one sum is possible because all the coefficients $t^{(i)}\pm t^{(j)}$ appearing in front of $\om$ take integer values.

Finally, since $\cos$ and $\sin$ have definite parity, one can reduce the Laurent series \Eq{laur} to a sum on $l\in\mathbb{N}$,
\ba
\Psi(f)&=&\sum_{l=0}^{+\infty}\left[(\tilde A^{\rm c}_l+\tilde A^{\rm c}_{-l})\cos(l\om\ln f)+(\tilde A^{\rm s}_l-\tilde A^{\rm s}_{-l})\sin(l\om\ln f)\right]\nonumber\\
&=& \sum_{l=0}^{+\infty} F_l(f)\,,
\ea
which is exactly the modulation factor in \Eq{master1}--\Eq{master2} with
\be
A^{\rm c}_l \coloneqq  \tilde A^{\rm c}_l+\tilde A^{\rm c}_{-l}\,,\qquad A^{\rm s}_l \coloneqq  \tilde A^{\rm s}_l-\tilde A^{\rm s}_{-l}\,,
\ee
and normalization $A^{\rm c}_0=1$. The upper limit of the sum is then set to some finite $l_{\rm max}$ either for numerical purposes or because the underlying theoretical model predicts $A^{\rm c,s}_l=0$ for $l>l_{\rm max}$.


\section{Analytic understanding for the parameter dependence of the Fisher errors}\label{appB}

Here, we present an analytic estimate to understand the tendency of the parameter dependence of the Fisher errors. By defining
\ba
\Omega_n^2(f) \coloneqq \left(\frac{10\pi^2} {3H_0^2}\right)^2 \frac{1}{2T_{\rm obs}}\sum_{(I,J)}\frac{f^6 S_I(f) S_J(f)}{ |\gamma_{IJ}(f)|^2} \,,
\ea
we can rewrite the SNR \eqref{SNR} and the Fisher matrix \eqref{Fisher} as 
\begin{eqnarray}
&& {\rm SNR}^2\ = \int^{+\infty}_0 \rmd f \frac{\Omega_{\textsc{gw}}^2(f)}{\Omega_n^2(f)}\,, \\
&&{\cal F}_{ij}=\int^{+\infty}_0 \rmd f \frac{\partial_{p_i}\Omega_{\textsc{gw}}(f) \partial_{p_j}\Omega_{\textsc{gw}}(f)}{\Omega_n^2(f)} \,.
\end{eqnarray}
Then, for the $l_{\rm max}=1$ case, we obtain
\begin{eqnarray}
&& {\cal F}_{\ln\Omega_0\ln\Omega_0} = \int \rmd f \frac{\Omega_{\textsc{gw}}^2(f)}{\Omega_n^2(f)} = {\rm SNR}^2\,, \\
&& {\cal F}_{n_{\rm t} n_{\rm t}} = \int \rmd f \frac{\Omega_{\textsc{gw}}^2(f)}{\Omega_n^2(f)} \ln^2\left(\frac{f}{f_*}\right) \,, \\
&& {\cal F}_{\omega\omega} \simeq \int \rmd f \frac{\Omega_{\textsc{gw}}^2(f)}{\Omega_n^2(f)} A_1^2 \ln^2 \left(\frac{f}{f_*}\right) \cos^2\left(\omega\ln{\frac{f}{f_*}}+\Phi_1\right)\,, \\
&& {\cal F}_{A_1 A_1} \simeq \int \rmd f \frac{\Omega_{\textsc{gw}}^2(f)}{\Omega_n^2(f)} \sin^2\left(\omega\ln{\frac{f}{f_*}}+\Phi_1\right)\,, \\
&& {\cal F}_{\phi_1 \phi_1} \simeq \int \rmd f \frac{\Omega_{\textsc{gw}}^2(f)}{\Omega_n^2(f)} A_1^2 \cos^2\left(\omega\ln{\frac{f}{f_*}}+\Phi_1\right)\,,
\end{eqnarray}
where we have used the approximation $A_1\ll 1$.

The non-marginalized error is given by $\theta_{i}\propto 1/\sqrt{\cF_{ii}}$ and it is useful to roughly estimate the parameter dependencies. First, it is straightforward to obtain 
\be
 \theta_{\ln \Omega_0} = \frac{1}{\rm SNR}\,.
\ee
The rest of the parameters require numerical computations for a given detector sensitivity and spectral shape. In fact, the integration for $\theta_{n_{\rm t}}$ includes only $\ln(f/f_*)$, so the detailed shape of ${\Omega_{\textsc{gw}}^2(f)}/{\Omega_n^2(f)}$ strongly influences the result. On the other hand, the tendency of the oscillation parameters can be understood relatively easily by assuming frequency-independence of ${\Omega_{\textsc{gw}}^2(f)}/{\Omega_n^2(f)}={\rm const}$. Considering that the integration of the oscillation part does not induce additional parameter dependence, we obtain
\begin{eqnarray}
&& \theta_{\ln \omega} = \frac{\theta_\om}{\om} \propto \frac{1}{A_1\om}\,, \\
&& \theta_{\ln A_1} = \frac{\theta_{A_1}}{A_1} \propto \frac{1}{A_1}\,, \\
&& \theta_{\phi}\propto \frac{1}{A_1}\,.
\end{eqnarray}
These results agree with the trends in the bottom panels of figure \ref{Fig2}.



\bigskip

\section*{References}

\begin{thebibliography}{99}
\bibitem{Abbott:2016blz} \au{B P}{Abbott} {\it et al} [LIGO Scientific and \textsc{Virgo} Collaborations]  2016 \tia{Observation of gravitational waves from a binary black hole merger} \doinn{10.1103/PhysRevLett.116.061102}{Phys.\ Rev.\ Lett.}{116}{061102}{2016} [\arX{1602.03837}]
\bibitem{TheLIGOScientific:2016src} \au{B P}{Abbott} {\it et al} [LIGO Scientific and \textsc{Virgo} Collaborations] 2016 \tia{Tests of general relativity with GW150914} \doinn{10.1103/PhysRevLett.116.221101}{Phys.\ Rev.\ Lett.}{116}{221101}{2016} [\arX{1602.03841}]

\hspace{-.35cm} \au{B P}{Abbott} {\it et al} [LIGO Scientific and \textsc{Virgo} Collaborations] 2018 \doinn{10.1103/PhysRevLett.121.129902}{Phys.\ Rev.\ Lett.}{121}{129902}{2018} (erratum)
\bibitem{Mirshekari:2011yq} \au{S}{Mirshekari}, \au{N}{Yunes} and \au{C M}{Will} 2012 \tia{Constraining generic Lorentz violation and the speed of the graviton with gravitational waves} \doin{10.1103/PhysRevD.85.024041}{Phys.\ Rev.}{D}{85}{024041}{2012} [\arX{1110.2720}]
\bibitem{Canizares:2012is} \au{P}{Canizares}, \au{J R}{Gair} and \au{C F}{Sopuerta} 2012 \tia{Testing Chern--Simons modified gravity with gravitational-wave detections of extreme-mass-ratio binaries} \doin{10.1103/PhysRevD.86.044010}{Phys.\ Rev.}{D}{86}{044010}{2012} [\arX{1205.1253}]
\bibitem{Ellis:2016rrr} \au{J}{Ellis}, \au{N E}{Mavromatos} and \au{D V}{Nanopoulos} 2016 \tia{Comments on graviton propagation in light of GW150914} \doin{10.1142/S0217732316750018}{Mod.\ Phys.\ Lett.}{A}{31}{1650155}{2016} [\arX{1602.04764}]
\bibitem{Yunes:2016jcc}   \au{N}{Yunes}, \au{K}{Yagi} and \au{F}{Pretorius} 2016 \tia{Theoretical physics implications of the binary black-hole merger GW150914} \doin{10.1103/PhysRevD.94.084002}{Phys.\ Rev.}{D}{94}{084002}{2016} [\arX{1603.08955}]
\bibitem{Arzano:2016twc} \au{M}{Arzano} and \au{G}{Calcagni} 2016 \tia{What gravity waves are telling about quantum spacetime} \doin{10.1103/PhysRevD.93.124065}{Phys.\ Rev.}{D}{93}{124065}{2016} [\arX{1604.00541}]
\bibitem{Calcagni:2019kzo} \au{G}{Calcagni}, \au{S}{Kuroyanagi}, \au{S}{Marsat}, \au{M}{Sakellariadou}, \au{N}{Tamanini} and \au{G}{Tasinato} 2019 \tia{Gravitational-wave luminosity distance in quantum gravity} \doin{10.1016/j.physletb.2019.135000}{Phys.\ Lett.}{B}{798}{135000}{2019} [\arX{1904.00384}]
\bibitem{Belgacem:2019pkk} \au{E}{Belgacem} {et al.} [LISA Cosmology Working Group] 2019 \tia{Testing modified gravity at cosmological distances with LISA standard sirens} \doij{10.1088/1475-7516/2019/07/024}{J.\ Cosmol.\ Astropart.\ Phys.}{JCAP07}{024}{2019} [\arX{1906.01593}]
\bibitem{Calcagni:2019ngc} \au{G}{Calcagni}, \au{S}{Kuroyanagi}, \au{S}{Marsat}, \au{M}{Sakellariadou}, \au{N}{Tamanini} and \au{G}{Tasinato} 2019 \tia{Quantum gravity and gravitational-wave astronomy} \doij{10.1088/1475-7516/2019/10/012}{J.\ Cosmol.\ Astropart.\ Phys.}{JCAP10}{012}{2019} [\arX{1907.02489}]
\bibitem{Belgacem:2020pdz} \au{E}{Belgacem}, \au{Y}{Dirian}, \au{A}{Finke}, \au{S}{Foffa} and \au{M}{Maggiore} 2020 \tia{Gravity in the infrared and effective nonlocal models} \doij{10.1088/1475-7516/2020/04/010}{J.\ Cosmol.\ Astropart.\ Phys.}{JCAP04}{010}{2020} [\arX{2001.07619}]
\bibitem{Barausse:2020rsu} \au{E}{Barausse} {et al.} [LISA Fundamental Physics Working Group] 2020 \tia{Prospects for fundamental physics with LISA} \doinn{10.1007/s10714-020-02691-1}{Gen.\ Relativ.\ Gravit.}{52}{81}{2020} [\arX{2001.09793}]
\bibitem{Calcagni:2020tvw} \au{G}{Calcagni} and \au{S}{Kuroyanagi} 2021 \tia{Stochastic gravitational-wave background in quantum gravity}, \doij{10.1088/1475-7516/2021/03/019}{J.\ Cosmol.\ Astropart.\ Phys.}{JCAP03}{019}{2021} [\arX{2012.00170}]
\bibitem{Addazi:2021xuf} \au{A}{Addazi} {et al.} 2022 \tia{Quantum gravity phenomenology at the dawn of the multi-messenger era -- A review} \doinn{10.1016/j.ppnp.2022.103948}{Prog.\ Part.\ Nucl.\ Phys.}{125}{103948}{2022} [\arX{2111.05659}]
\bibitem{LISACosmologyWorkingGroup:2022wjo} \au{T}{Baker} {et al.} [LISA Cosmology Working Group] 2022 \tia{Measuring the propagation speed of gravitational waves with LISA} \doij{10.1088/1475-7516/2022/08/031}{J.\ Cosmol.\ Astropart.\ Phys.}{JCAP08}{031}{2022} [\arX{2203.00566}]
\bibitem{LISACosmologyWorkingGroup:2022jok} \au{P}{Auclair} {et al.} [LISA Cosmology Working Group] 2023 \tia{Cosmology with the Laser Interferometer Space Antenna} \doinn{10.1007/s41114-023-00045-2}{Living Rev.\ Relativ.}{26}{5}{2023} [\arX{2204.05434}]
\bibitem{LISA:2022kgy} \au{K G}{Arun} {et al.} [LISA Fundamental Physics Working Group] 2022 \tia{New horizons for fundamental physics with LISA} \doinn{10.1007/s41114-022-00036-9}{Living Rev.\ Relativ.}{25}{4}{2022} [\arX{2205.01597}]
\bibitem{Christensen:2018iqi} \au{N}{Christensen} 2019 \tia{Stochastic gravitational wave backgrounds} \doinn{10.1088/1361-6633/aae6b5}{Rept.\ Prog.\ Phys.}{82}{016903}{2019} [\arX{1811.08797}]
\bibitem{Renzini:2022alw} \au{A I}{Renzini}, \au{B}{Goncharov}, \au{A C}{Jenkins} and \au{P M}{Meyers} 2022 \tia{Stochastic gravitational-wave backgrounds: current detection efforts and future prospects} \doinn{10.3390/galaxies10010034}{Galaxies}{10}{34}{2022} [\arX{2202.00178}]
\bibitem{vanRemortel:2022fkb} \au{N}{van Remortel}, \au{K}{Janssens} and \au{K}{Turbang} 2023 \tia{Stochastic gravitational wave background: methods and implications} \doinn{10.1016/j.ppnp.2022.104003}{Prog.\ Part.\ Nucl.\ Phys.}{128}{104003}{2023} [\arX{2210.00761}]
\bibitem{NANOGrav:2023gor} \au{G}{Agazie} {et al.} [NANOGrav] 2023 \tia{The NANOGrav 15-year data set: evidence for a gravitational-wave background} \doinn{10.3847/2041-8213/acdac6}{Astrophys.\ J.\ Lett.}{951}{L8}{2023} [\arX{2306.16213}]
\bibitem{Antoniadis:2023ott} \au{J}{Antoniadis} {et al.} 2023 \tia{The second data release from the European Pulsar Timing Array III. Search for gravitational wave signals} \doinn{10.1051/0004-6361/202346844}{Astron.\ Astrophys.}{678}{A50}{2023} [\arX{2306.16214}]
\bibitem{Reardon:2023gzh} \au{D J}{Reardon} {et al.} [PPTA] 2023 \tia{Search for an isotropic gravitational-wave background with the Parkes Pulsar Timing Array} \doinn{10.3847/2041-8213/acdd02}{Astrophys.\ J. Lett.}{951}{L6}{2023} [\arX{2306.16215}]
\bibitem{Xu:2023wog} \au{H}{Xu} {et al.} [CPTA] 2023 \tia{Searching for the nano-Hertz stochastic gravitational wave background with the Chinese Pulsar Timing Array Data Release I} \doinn{10.1088/1674-4527/acdfa5}{Res.\ Astron.\ Astrophys.}{23}{075024}{2023} [\arX{2306.16216}]
\bibitem{Calcagni:2013qqa}  \au{G}{Calcagni} and \au{G}{Nardelli} 2014 \tia{Quantum field theory with varying couplings} \doin{10.1142/S0217751X14500122}{Int.\ J.\ Mod.\ Phys.}{A}{29}{1450012}{2014} [\arX{1306.0629}]
\bibitem{Calcagni:2016azd} \au{G}{Calcagni} 2017 \tia{Multifractional theories: an unconventional review} \doij{10.1007/JHEP03(2017)138}{J.\ High Energy Phys.}{JHEP03}{138}{2017} [\arX{1612.05632}]
\bibitem{Calcagni:2021ipd} \au{G}{Calcagni} 2021 \tia{Multifractional theories: an updated review} \doin{10.1142/S021773232140006X}{Mod.\ Phys.\ Lett.}{A}{36}{2140006}{2021} [\arX{2103.06557}]
\bibitem{Calcagni:2017via} \au{G}{Calcagni} 2017 \tia{Complex dimensions and their observability} \doin{10.1103/PhysRevD.96.046001}{Phys.\ Rev.}{D}{96}{046001}{2017} [\arX{1705.01619}]
\bibitem{Maggiore:2019uih} \au{M}{Maggiore} {\it et al} 
 2020 \tia{Science case for the Einstein Telescope} \doij{10.1088/1475-7516/2020/03/050}{J.\ Cosmol.\ Astropart.\ Phys.}{JCAP03}{050}{2020} [\arX{1912.02622}]
\bibitem{Tom67} \au{K}{Tomita} 1967 \tia{Non-linear theory of gravitational instability in the expanding universe} \doinn{10.1143/PTP.37.831}{Prog.\ Theor.\ Phys.}{37}{831}{1967}
\bibitem{Matarrese:1992rp} \au{S}{Matarrese}, \au{O}{Pantano} and \au{D}{Saez} 1993 \tia{General-relativistic approach to the nonlinear evolution of collisionless matter} \doin{10.1103/PhysRevD.47.1311}{Phys.\ Rev.}{D}{47}{1311}{1993}
\bibitem{Matarrese:1993zf} \au{S}{Matarrese}, \au{O}{Pantano} and \au{D}{Saez} 1994 \tia{General relativistic dynamics of irrotational dust: cosmological implications} \doinn{10.1103/PhysRevLett.72.320}{Phys.\ Rev.\ Lett.}{72}{320}{1994} [\oarX{astro-ph/9310036}]
\bibitem{Matarrese:1997ay} \au{S}{Matarrese}, \au{S}{Mollerach} and \au{M}{Bruni} 1998 \tia{Relativistic second-order perturbations of the Einstein--de Sitter universe} \doin{10.1103/PhysRevD.58.043504}{Phys.\ Rev.}{D}{58}{043504}{1998} [\oarX{astro-ph/9707278}]
\bibitem{Mollerach:2003nq} \au{S}{Mollerach}, \au{D}{Harari} and \au{S}{Matarrese} 2004 \tia{CMB polarization from secondary vector and tensor modes} \doin{10.1103/PhysRevD.69.063002}{Phys.\ Rev.}{D}{69}{063002}{2004} [\oarX{astro-ph/0310711}]
\bibitem{Ananda:2006af} \au{K N}{Ananda}, \au{C}{Clarkson} and \au{D}{Wands} 2007 \tia{The cosmological gravitational wave background from primordial density perturbations} \doin{10.1103/PhysRevD.75.123518}{Phys.\ Rev.}{D}{75}{123518}{2007} [\oarX{gr-qc/0612013}]
\bibitem{Baumann:2007zm} \au{D}{Baumann}, \au{P J}{Steinhardt}, \au{K}{Takahashi} and \au{K}{Ichiki} 2007 \tia{Gravitational wave spectrum induced by primordial scalar perturbations} \doin{10.1103/PhysRevD.76.084019}{Phys.\ Rev.}{D}{76}{084019}{2007} [\oarX{hep-th/0703290}]
\bibitem{Fumagalli:2021cel} \au{J}{Fumagalli}, \au{S}{Renaux-Petel} and \au{L T}{Witkowski} 2021 \tia{Resonant features in the stochastic gravitational wave background} \doij{10.1088/1475-7516/2021/08/059}{J.\ Cosmol.\ Astropart.\ Phys.}{JCAP08}{059}{2021} [\arX{2105.06481}]
\bibitem{Chen:2008wn} \au{X}{Chen}, \au{R}{Easther} and \au{E A}{Lim} 2008 \tia{Generation and characterization of large non-Gaussianities in single field inflation} \doij{10.1088/1475-7516/2008/04/010}{J.\ Cosmol.\ Astropart.\ Phys.}{JCAP04}{010}{2008} [\arX{0801.3295}]
\bibitem{Achucarro:2010jv} \au{A}{Ach\'ucarro}, \au{J O}{Gong}, \au{S}{Hardeman}, \au{G A}{Palma} and \au{S P}{Patil} 2011 \tia{Mass hierarchies and non-decoupling in multi-scalar field dynamics} \doin{10.1103/PhysRevD.84.043502}{Phys.\ Rev.}{D}{84}{043502}{2011} [\arX{1005.3848}]
\bibitem{Achucarro:2010da} \au{A}{Ach\'ucarro}, \au{J O}{Gong}, \au{S}{Hardeman}, \au{G A}{Palma} and \au{S P}{Patil} 2011 \tia{Features of heavy physics in the CMB power spectrum} \doij{10.1088/1475-7516/2011/01/030}{J.\ Cosmol.\ Astropart.\ Phys.}{JCAP01}{030}{2011} [\arX{1010.3693}]
\bibitem{Chen:2011zf} \au{X}{Chen} 2012 \tia{Primordial features as evidence for inflation} \doij{10.1088/1475-7516/2012/01/038}{J.\ Cosmol.\ Astropart.\ Phys.}{JCAP01}{038}{2012} [\arX{1104.1323}]
\bibitem{Chen:2011tu} \au{X}{Chen} 2011 \tia{Fingerprints of primordial universe paradigms as features in density perturbations} \doin{10.1016/j.physletb.2011.11.009}{Phys.\ Lett.}{B}{706}{111}{2011} [\arX{1106.1635}]
\bibitem{Shiu:2011qw} \au{G}{Shiu} and \au{J}{Xu} 2011 \tia{Effective field theory and decoupling in multi-field inflation: an illustrative case study} \doin{10.1103/PhysRevD.84.103509}{Phys.\ Rev.}{D}{84}{103509}{2011} [\arX{1108.0981}]
\bibitem{Gao:2012uq} \au{X}{Gao}, \au{D}{Langlois} and \au{S}{Mizuno} 2012 \tia{Influence of heavy modes on perturbations in multiple field inflation} \doij{10.1088/1475-7516/2012/10/040}{J.\ Cosmol.\ Astropart.\ Phys.}{JCAP10}{040}{2012} [\arX{1205.5275}]
\bibitem{Gao:2013ota} \au{X}{Gao}, \au{D}{Langlois} and \au{S}{Mizuno} 2013 \tia{Oscillatory features in the curvature power spectrum after a sudden turn of the inflationary trajectory} \doij{10.1088/1475-7516/2013/10/023}{J.\ Cosmol.\ Astropart.\ Phys.}{JCAP10}{023}{2013} [\arX{1306.5680}]
\bibitem{Noumi:2013cfa} \au{T}{Noumi} and \au{M}{Yamaguchi} 2013 \tia{Primordial spectra from sudden turning trajectory}, \doij{10.1088/1475-7516/2013/12/038}{J.\ Cosmol.\ Astropart.\ Phys.}{JCAP12}{038}{2013} [\arX{1307.7110}]
\bibitem{Chen:2014joa} \au{X}{Chen} and \au{M H}{Namjoo} 2014 \tia{Standard clock in primordial density perturbations and cosmic microwave background} \doin{10.1016/j.physletb.2014.11.002}{Phys.\ Lett.}{B}{739}{285}{2014} [\arX{1404.1536}]
\bibitem{Chen:2014cwa} \au{X}{Chen}, \au{M H}{Namjoo} and \au{Y}{Wang} 2015 \tia{Models of the primordial standard clock} \doij{10.1088/1475-7516/2015/02/027}{J.\ Cosmol.\ Astropart.\ Phys.}{JCAP02}{027}{2015} [\arX{1411.2349}]
\bibitem{Braglia:2020taf} \au{M}{Braglia}, \au{X}{Chen} and \au{D K}{Hazra} 2021 \tia{Probing primordial features with the stochastic gravitational wave background} \doij{10.1088/1475-7516/2021/03/005}{J.\ Cosmol.\ Astropart.\ Phys.}{JCAP03}{005}{2021} [\arX{2012.05821}]
\bibitem{Flauger:2009ab} \au{R}{Flauger}, \au{L}{McAllister}, \au{E}{Pajer}, \au{A}{Westphal} and \au{G}{Xu} 2010 \tia{Oscillations in the CMB from axion monodromy inflation} \doij{10.1088/1475-7516/2010/06/009}{J.\ Cosmol.\ Astropart.\ Phys.}{JCAP06}{009}{2010} [\arX{0907.2916}]
\bibitem{Flauger:2014ana} \au{R}{Flauger}, \au{L}{McAllister}, \au{E}{Silverstein} and \au{A}{Westphal} 2017 \tia{Drifting oscillations in axion monodromy} \doij{10.1088/1475-7516/2017/10/055}{J.\ Cosmol.\ Astropart.\ Phys.}{JCAP10}{055}{2017} [\arX{1412.1814}]
\bibitem{Behbahani:2011it} \au{S.R}{Behbahani}, \au{A}{Dymarsky}, \au{M}{Mirbabayi} and \au{L}{Senatore} 2012 \tia{(Small) resonant non-Gaussianities: signatures of a discrete shift symmetry in the effective field theory of inflation} \doij{10.1088/1475-7516/2012/12/036}{J.\ Cosmol.\ Astropart.\ Phys.}{JCAP12}{036}{2012} [\arX{1111.3373}]
\bibitem{Bhattacharya:2022fze} \au{S}{Bhattacharya} and \au{I}{Zavala} 2023 \tia{Sharp turns in axion monodromy: primordial black holes and gravitational waves} \doij{10.1088/1475-7516/2023/04/065}{J.\ Cosmol.\ Astropart.\ Phys.}{JCAP04}{065}{2023} [\arX{2205.06065}]
\bibitem{Giddings:2001yu} \au{S B}{Giddings}, \au{S}{Kachru} and \au{J}{Polchinski} 2002 \tia{Hierarchies from fluxes in string compactifications} \doin{10.1103/PhysRevD.66.106006}{Phys.\ Rev.}{D}{66}{106006}{2002} [\arX{hep-th/0105097}]
\bibitem{Mavromatos:2022yql} \au{N E}{Mavromatos}, \au{V C}{Spanos} and \au{I D}{Stamou} 2022 \tia{Primordial black holes and gravitational waves in multiaxion-Chern--Simons inflation} \doin{10.1103/PhysRevD.106.063532}{Phys.\ Rev.}{D}{106}{063532}{2022} [\arX{2206.07963}]
\bibitem{Brattan:2017yzx} \au{D K}{Brattan}, \au{O}{Ovdat} and \au{E}{Akkermans} 2018 \tia{Scale anomaly of a Lifshitz scalar: a universal quantum phase transition to discrete scale invariance} \doin{10.1103/PhysRevD.97.061701}{Phys.\ Rev.}{D}{97}{061701}{2018} [\arX{1706.00016}]
\bibitem{Turner:1993vb} \au{M S}{Turner}, \au{M J}{White} and \au{J E}{Lidsey} 1993 \tia{Tensor perturbations in inflationary models as a probe of cosmology} \doin{10.1103/PhysRevD.48.4613}{Phys.\ Rev.}{D}{48}{4613}{1993} [\oarX{astro-ph/9306029}]
\bibitem{Kuroyanagi:2008ye} \au{S}{Kuroyanagi}, \au{T}{Chiba} and \au{N}{Sugiyama} 2009 \tia{Precision calculations of the gravitational wave background spectrum from inflation} \doin{10.1103/PhysRevD.79.103501}{Phys.\ Rev.}{D}{79}{103501}{2009} [\arX{0804.3249}]
\bibitem{Nakayama:2008wy} \au{K}{Nakayama}, \au{S}{Saito}, \au{Y}{Suwa} and \au{J}{Yokoyama} 2008 \tia{Probing reheating temperature of the universe with gravitational wave background} \doij{10.1088/1475-7516/2008/06/020}{J.\ Cosmol.\ Astropart.\ Phys.}{JCAP06}{020}{2008} [\arX{0804.1827}]
\bibitem{Kuroyanagi:2014nba} \au{S}{Kuroyanagi}, \au{T}{Takahashi} and \au{S}{Yokoyama} 2015 \tia{Blue-tilted tensor spectrum and thermal history of the universe} \doij{10.1088/1475-7516/2015/02/003}{J.\ Cosmol.\ Astropart.\ Phys.}{JCAP02}{003}{2015} [\arX{1407.4785}]
\bibitem{Sornette:1997pb} \au{D}{Sornette} 1998 \tia{Discrete scale invariance and complex dimensions} \doinn{10.1016/S0370-1573(97)00076-8}{Phys.\ Rept.}{297}{239}{1998} [\oarX{cond-mat/9707012}]
\bibitem{BGM1} \au{D}{Bessis}, \au{J S}{Geronimo} and \au{P}{Moussa} 1983 \tia{Complex spectral dimensionality on fractal structures} \doinn{10.1051/jphyslet:019830044024097700}{J.\ Phys.\ Lett.\ (Paris)}{44}{977}{1983}
\bibitem{DDSI}  \au{B}{Derrida}, \au{L}{De Seze} and \au{C}{Itzykson} 1983 \tia{Fractal structure of zeros in hierarchical models} \doinn{10.1007/BF01018834}{J.\ Stat.\ Phys.}{33}{559}{1983}
\bibitem{Bessis:1983nt} \au{D}{Bessis}, \au{J S}{Geronimo} and \au{P}{Moussa} 1984 \tia{Mellin transforms associated with Julia sets and physical applications} \doinn{10.1007/BF01770350}{J.\ Stat.\ Phys.}{34}{75}{1984}
\bibitem{NLM}   \au{R R}{Nigmatullin} and \au{A}{Le M\'ehaut\'e} 2005 \tia{Is there geometrical/physical meaning of the fractional integral with complex exponent?} \doinn{10.1016/j.jnoncrysol.2005.05.035}{J.\ Non-Cryst.\ Solids}{351}{2888}{2005}
\bibitem{LvF}   \au{M L}{Lapidus} and \au{M}{van Frankenhuysen} 2006 \book{Fractal Geometry, Complex Dimensions and Zeta Functions}{Springer}{New York}{U.S.A.}{2006}
\bibitem{Calcagni:2016edi} \au{G}{Calcagni} 2016 \tia{ABC of multi-fractal spacetimes and fractional sea turtles} \doin{10.1140/epjc/s10052-016-4021-0}{Eur.\ Phys.\ J.}{C}{76}{181}{2016} [\arX{1602.01470}]
\bibitem{Crane:1985ex} \au{L}{Crane} and \au{L}{Smolin} 1985 \tia{Space-time foam as the universal regulator} \doinn{10.1007/BF00773626}{Gen.\ Relativ.\ Gravit.}{17}{1209}{1985}
\bibitem{tHooft:1993dmi}  \au{G}{'t Hooft} 1993 \tia{Dimensional reduction in quantum gravity} in \procsinm{Salamfestschrift}{\ua{A}{Ali}, \ua{J}{Ellis} and \ua{S}{Randjbar-Daemi}}{World Scientific}{Singapore}{1993} [{\it Conf.\ Proc.} C {930308} (1993) 284] [\oarX{gr-qc/9310026}]
\bibitem{Ambjorn:2005db} \au{J}{Ambj{\o}rn}, \au{J}{Jurkiewicz} and \au{R}{Loll} 2005 \tia{Spectral dimension of the universe} \doinn{10.1103/PhysRevLett.95.171301}{Phys.\ Rev.\ Lett.}{95}{171301}{2005} [\oarX{hep-th/0505113}]
\bibitem{Lauscher:2005qz} \au{O}{Lauscher} and \au{M}{Reuter} 2005 \tia{Fractal spacetime structure in asymptotically safe gravity} \doij{10.1088/1126-6708/2005/10/050}{J.\ High Energy Phys.}{JHEP10}{050}{2005} [\oarX{hep-th/0508202}]
\bibitem{Benedetti:2008gu} \au{D}{Benedetti} 2009 \tia{Fractal properties of quantum spacetime} \doinn{10.1103/PhysRevLett.102.111303}{Phys.\ Rev.\ Lett.}{102}{111303}{2009} [\arX{0811.1396}]
\bibitem{Carlip:2009kf} \au{S}{Carlip} 2009 \tia{Spontaneous dimensional reduction in short-distance quantum gravity?} \doinn{10.1063/1.3284402}{AIP Conf.\ Proc.}{1196}{72}{2009} [\arX{0909.3329}]
\bibitem{Calcagni:2009kc}  \au{G}{Calcagni} 2010 \tia{Fractal universe and quantum gravity} \doinn{10.1103/PhysRevLett.104.251301}{Phys.\ Rev.\ Lett.}{104}{251301}{2010} [\arX{0912.3142}]
\bibitem{Calcagni:2016xtk} \au{G}{Calcagni} 2017 \tia{Multiscale spacetimes from first principles} \doin{10.1103/PhysRevD.95.064057}{Phys.\ Rev.}{D}{95}{064057}{2017} [\arX{1609.02776}]
\bibitem{Carlip:2017eud} \au{S}{Carlip} 2017 \tia{Dimension and dimensional reduction in quantum gravity} \doinn{10.1088/1361-6382/aa8535}{Classical Quantum Gravity}{34}{193001}{2017} [\arX{1705.05417}]
\bibitem{Mielczarek:2017cdp}  \au{J}{Mielczarek} and \au{T}{Trześniewski} 2018 \tia{Towards the map of quantum gravity} \doinn{10.1007/s10714-018-2391-3}{Gen.\ Relativ.\ Gravit.}{50}{68}{2018} [\arX{1708.07445}]
\bibitem{Carlip:2019onx} \au{S}{Carlip} 2019 \tia{Dimension and dimensional reduction in quantum gravity} \doinn{10.3390/universe5030083}{Universe}{5}{83}{2019} [\arX{1904.04379}]
\bibitem{Calcagni:2014cza} \au{G}{Calcagni}, \au{D}{Oriti} and \au{J}{Th\"urigen} 2015 \tia{Dimensional flow in discrete quantum geometries} \doin{10.1103/PhysRevD.91.084047}{Phys.\ Rev.}{D}{91}{084047}{2015} [\arX{1412.8390}]
\bibitem{Jercher:2023rno} \au{A}{Jercher}, \au{S}{Steinhaus} and \au{J}{Th\"urigen} 2023 \tia{Curvature effects in the spectral dimension of spin foams} \doin{10.1103/PhysRevD.108.066011}{Phys.\ Rev.}{D}{108}{066011}{2023} [\arX{2304.13058}]
\bibitem{Banks:2010zn} \au{T}{Banks} and \au{N}{Seiberg} 2011 \tia{Symmetries and strings in field theory and gravity}, \doin{10.1103/PhysRevD.83.084019}{Phys.\ Rev.}{D}{83}{084019}{2011} [\arX{1011.5120}]
\bibitem{Harlow:2018jwu} \au{D}{Harlow} and \au{H}{Ooguri} 2019 \tia{Constraints on symmetries from holography} \doinn{10.1103/PhysRevLett.122.191601}{Phys.\ Rev.\ Lett.}{122}{191601}{2019} [\arX{1810.05337}]
\bibitem{Harlow:2018tng} \au{D}{Harlow} and \au{H}{Ooguri} 2021 \tia{Symmetries in quantum field theory and quantum gravity} \doinn{10.1007/s00220-021-04040-y}{Commun.\ Math.\ Phys.}{383}{1669}{2021} [\arX{1810.05338}]
\bibitem{Steinhaus:2018aav} \au{S}{Steinhaus} and \au{J}{Th\"urigen} 2018 \tia{Emergence of spacetime in a restricted spin-foam model} \doin{10.1103/PhysRevD.98.026013}{Phys.\ Rev.}{D}{98}{026013}{2018} [\arX{1803.10289}]
\bibitem{Thu23} \au{J}{Th\"urigen} 2023 private communication
\bibitem{Martin:2000xs} \au{J}{Martin} and \au{R H}{Brandenberger} 2001 \tia{The trans-Planckian problem of inflationary cosmology} \doin{10.1103/PhysRevD.63.123501}{Phys.\ Rev.}{D}{63}{123501}{2001} [\oarX{hep-th/0005209}]
\bibitem{Easther:2001fz} \au{R}{Easther}, \au{B R}{Greene}, \au{W H}{Kinney} and \au{G}{Shiu} 2003 \tia{Imprints of short distance physics on inflationary cosmology} \doin{10.1103/PhysRevD.67.063508}{Phys.\ Rev.}{D}{67}{063508}{2003} [\oarX{hep-th/0110226}]
\bibitem{Easther:2001fi} \au{R}{Easther}, \au{B R}{Greene}, \au{W H}{Kinney} and \au{G}{Shiu} 2001 \tia{Inflation as a probe of short distance physics} \doin{10.1103/PhysRevD.64.103502}{Phys.\ Rev.}{D}{64}{103502}{2001} [\oarX{hep-th/0104102}]
\bibitem{Easther:2002xe} \au{R}{Easther}, \au{B R}{Greene}, \au{W H}{Kinney} and \au{G}{Shiu} 2002 \tia{Generic estimate of trans-Planckian modifications to the primordial power spectrum in inflation} \doin{10.1103/PhysRevD.66.023518}{Phys.\ Rev.}{D}{66}{023518}{2002} [\oarX{hep-th/0204129}]
\bibitem{Martin:2003kp} \au{J}{Martin} and \au{R H}{Brandenberger} 2003 \tia{On the dependence of the spectra of fluctuations in inflationary cosmology on trans-Planckian physics} \doin{10.1103/PhysRevD.68.063513}{Phys.\ Rev.}{D}{68}{063513}{2003} [\oarX{hep-th/0305161}]
\bibitem{Greene:2004fln} \au{B}{Greene}, \au{K}{Schalm}, \au{J P}{van der Schaar} and \au{G}{Shiu} 2005 \tia{Extracting new physics from the CMB} \oarX{astro-ph/0503458}
\bibitem{Chen:2010bka} \au{X}{Chen} 2010 \tia{Folded resonant non-Gaussianity in general single field inflation} \doij{10.1088/1475-7516/2010/12/003}{J.\ Cosmol.\ Astropart.\ Phys.}{JCAP12}{003}{2010} [\arX{1008.2485}]
\bibitem{Danielsson:2002kx} \au{U H}{Danielsson} 2002 \tia{Note on inflation and trans-Planckian physics} \doin{10.1103/PhysRevD.66.023511}{Phys.\ Rev.}{D}{66}{023511}{2002} [\oarX{hep-th/0203198}]
\bibitem{Bozza:2003pr} \au{V}{Bozza}, \au{M}{Giovannini} and \au{G}{Veneziano} 2003 \tia{Cosmological perturbations from a new physics hypersurface} \doij{10.1088/1475-7516/2003/05/001}{J.\ Cosmol.\ Astropart.\ Phys.}{JCAP05}{001}{2003} [\oarX{hep-th/0302184}]
\bibitem{Jackson:2010cw} \au{M G}{Jackson} and \au{K}{Schalm} 2012 \tia{Model independent signatures of new physics in the inflationary power spectrum} \doinn{10.1103/PhysRevLett.108.111301}{Phys.\ Rev.\ Lett.}{108}{111301}{2012} [\arX{1007.0185}]
\bibitem{Jackson:2013mka} \au{M G}{Jackson}, \au{B}{Wandelt} and \au{F}{Bouchet} 2014 \tia{Angular correlation functions for models with logarithmic oscillations} \doin{10.1103/PhysRevD.89.023510}{Phys.\ Rev.}{D}{89}{023510}{2014} [\arX{1303.3499}]
\bibitem{Calcagni:2013yqa} \au{G}{Calcagni} 2013 \tia{Multi-scale gravity and cosmology} \doij{10.1088/1475-7516/2013/12/041}{J.\ Cosmol.\ Astropart.\ Phys.}{JCAP12}{041}{2013} [\arX{1307.6382}]
\bibitem{Calcagni:2016ofu} \au{G}{Calcagni}, \au{S}{Kuroyanagi} and \au{S}{Tsujikawa} 2016 \tia{Cosmic microwave background and inflation in multi-fractional spacetimes} \doij{10.1088/1475-7516/2016/08/039}{J.\ Cosmol.\ Astropart.\ Phys.}{JCAP08}{039}{2016} [\arX{1606.08449}]
\bibitem{Braglia:2021rej} \au{M}{Braglia}, \au{X}{Chen} and \au{D K}{Hazra} 2022 \tia{Primordial standard clock models and CMB residual anomalies} \doin{10.1103/PhysRevD.105.103523}{Phys.\ Rev.}{D}{105}{103523}{2022} [\arX{2108.10110}]
\bibitem{Allen:1997ad} \au{B}{Allen} and \au{J D}{Romano} 1999 \tia{Detecting a stochastic background of gravitational radiation: signal processing strategies and sensitivities} \doin{10.1103/PhysRevD.59.102001}{Phys.\ Rev.}{D}{59}{102001}{1999} [\oarX{gr-qc/9710117}]
\bibitem{Planck:2018jri} \au{Y}{Akrami} {\it et al} [\textsc{Planck} Collaboration] 2020 \tia{Planck 2018 results. X. Constraints on inflation} \doinn{10.1051/0004-6361/201833887}{Astron.\ Astrophys.}{641}{A10}{2020} [\arX{1807.06211}]
\bibitem{Seto:2005qy} \au{N}{Seto} 1006 \tia{Correlation analysis of stochastic gravitational wave background around 0.1-1 Hz} \doin{10.1103/PhysRevD.73.063001}{Phys.\ Rev.}{D}{73}{063001}{2006} [\oarX{gr-qc/0510067}]
\bibitem{Kuroyanagi:2009br} \au{S}{Kuroyanagi}, \au{C}{Gordon}, \au{J}{Silk} and \au{N}{Sugiyama} 2010 \tia{Forecast constraints on inflation from combined CMB and gravitational wave direct detection experiments} \doin{10.1103/PhysRevD.81.083524}{Phys.\ Rev.}{D}{81}{083524}{2010} [\arX{0912.3683}]

\hspace{-.35cm} \au{S}{Kuroyanagi}, \au{C}{Gordon}, \au{J}{Silk} and \au{N}{Sugiyama} 2010 \doin{10.1103/PhysRevD.82.069901}{Phys.\ Rev.}{D}{82}{069901}{2010} (erratum)
\bibitem{Branchesi:2023mws} \au{M}{Branchesi} {\it et al} 2023 \tia{Science with the Einstein Telescope: a comparison of different designs} 
\doij{10.1088/1475-7516/2023/07/068}{J.\ Cosmol.\ Astropart.\ Phys.}{JCAP07}{068}{2023} [\arX{2303.15923}]
\bibitem{ETsensitivity} \href{https://www.et-gw.eu/index.php/etsensitivities}{\cob https://www.et-gw.eu/index.php/etsensitivities}
\bibitem{Regimbau:2012ir} \au{T}{Regimbau} {et al.} 2012 \tia{Mock data challenge for the Einstein Gravitational-Wave Telescope} \doin{10.1103/PhysRevD.86.122001}{Phys.\ Rev.}{D}{86}{122001}{2012} [\arX{1201.3563}]
\bibitem{KAGRA:2021kbb} \au{R}{Abbott} {\it et al} [LIGO-Virgo-KAGRA] 2021 \tia{Upper limits on the isotropic gravitational-wave background from Advanced LIGO and Advanced Virgo's third observing run} \doin{10.1103/PhysRevD.104.022004}{Phys.\ Rev.}{D}{104}{022004}{2021} [\arX{2101.12130}]
\bibitem{Flauger:2020qyi} \au{R}{Flauger}, \au{N}{Karnesis}, \au{G}{Nardini}, \au{M}{Pieroni}, \au{A}{Ricciardone} and \au{J}{Torrado} 2021 \tia{Improved reconstruction of a stochastic gravitational wave background with LISA} \doij{10.1088/1475-7516/2021/01/059}{J.\ Cosmol.\ Astropart.\ Phys.}{JCAP01}{059}{2021} [\arX{2009.11845}]
\bibitem{BICEP:2021xfz} \au{P.A.R}{Ade} {\it et al.} [BICEP/Keck Collaboration], \tia{Improved constraints on primordial gravitational waves using Planck, WMAP, and BICEP/Keck observations through the 2018 observing season} \doinn{10.1103/PhysRevLett.127.151301}{Phys.\ Rev.\ Lett.}{127}{151301}{2021} [\arX{2110.00483}]

\bibitem{Adams:2001vc} \au{J A}{Adams}, \au{B}{Cresswell} and \au{R}{Easther} 2001 \tia{Inflationary perturbations from a potential with a step} \doin{10.1103/PhysRevD.64.123514}{Phys.\ Rev.}{D}{64}{123514}{2001} [\oarX{astro-ph/0102236}]
\bibitem{Bean:2008na} \au{R}{Bean}, \au{X}{Chen}, \au{G}{Hailu}, \au{S H H}{Tye} and \au{J}{Xu} 2008 \tia{Duality cascade in brane inflation} \doij{10.1088/1475-7516/2008/03/026}{J.\ Cosmol.\ Astropart.\ Phys.}{JCAP03}{026}{2008} [\arX{0802.0491}]
\bibitem{Park:2012rh} \au{M}{Park} and \au{L}{Sorbo} 2012 \tia{Sudden variations in the speed of sound during inflation: features in the power spectrum and bispectrum} \doin{10.1103/PhysRevD.85.083520}{Phys.\ Rev.}{D}{85}{083520}{2012} [\arX{1201.2903}] 
\bibitem{Miranda:2012rm} \au{V}{Miranda}, \au{W}{Hu} and \au{P}{Adshead} 2012 \tia{Warp features in DBI inflation} \doin{10.1103/PhysRevD.86.063529}{Phys.\ Rev.}{D}{2286}{063529}{2012} [\arX{1207.2186}]
\bibitem{Bartolo:2013exa} \au{N}{Bartolo}, \au{D}{Cannone} and \au{S}{Matarrese} 2013 \tia{The effective field theory of inflation models with sharp features} \doij{10.1088/1475-7516/2013/10/038}{J.\ Cosmol.\ Astropart.\ Phys.}{JCAP10}{038}{2013} [\arX{1307.3483}]
\bibitem{Hazra:2014jka} \au{D K}{Hazra}, \au{A}{Shafieloo}, \au{G F}{Smoot} and \au{A A}{Starobinsky} 2014 \tia{Inflation with whip-shaped suppressed scalar power spectra} \doinn{10.1103/PhysRevLett.113.071301}{Phys.\ Rev.\ Lett.}{113}{071301}{2014} [\arX{1404.0360}]
\bibitem{Hazra:2014goa} \au{D K}{Hazra}, \au{A}{Shafieloo}, \au{G F}{Smoot} and \au{A A}{Starobinsky} 2014 \tia{Wiggly whipped inflation} \doij{10.1088/1475-7516/2014/08/048}{J.\ Cosmol.\ Astropart.\ Phys.}{JCAP08}{048}{2014} [\arX{1405.2012}]
\bibitem{Palma:2014hra} \au{G A}{Palma} 2015 \tia{Untangling features in the primordial spectra} \doij{10.1088/1475-7516/2015/04/035}{J.\ Cosmol.\ Astropart.\ Phys.}{JCAP04}{035}{2015} [\arX{1412.5615}]
\bibitem{Hazra:2016fkm} \au{D K}{Hazra}, \au{A}{Shafieloo}, \au{G F}{Smoot} and \au{A A}{Starobinsky} 2016 \tia{Primordial features and Planck polarization} \doij{10.1088/1475-7516/2016/09/009}{J.\ Cosmol.\ Astropart.\ Phys.}{JCAP09}{009}{2016} [\arX{1605.02106}]
\bibitem{Ballesteros:2018wlw} \au{G}{Ballesteros}, \au{J}{Beltr\'an Jim\'enez} and \au{M}{Pieroni} 2019 \tia{Black hole formation from a general quadratic action for inflationary primordial fluctuations} \doij{10.1088/1475-7516/2019/06/016}{J.\ Cosmol.\ Astropart.\ Phys.}{JCAP06}{016}{2019} [\arX{1811.03065}]
\bibitem{Palma:2020ejf} \au{G A}{Palma}, \au{S}{Sypsas} and \au{C}{Zenteno} 2020 \tia{Seeding primordial black holes in multifield inflation} \doinn{10.1103/PhysRevLett.125.121301}{Phys.\ Rev.\ Lett.}{125}{121301}{2020} [\arX{2004.06106}]
\bibitem{Kefala:2020xsx} \au{K}{Kefala}, \au{G P}{Kodaxis}, \au{I D}{Stamou} and \au{N}{Tetradis} 2021 \tia{Features of the inflaton potential and the power spectrum of cosmological perturbations} \doin{10.1103/PhysRevD.104.023506}{Phys.\ Rev.}{D}{104}{023506}{2021} [\arX{2010.12483}]
\bibitem{Fumagalli:2020nvq} \au{J}{Fumagalli}, \au{S}{Renaux-Petel} and \au{L T}{Witkowski} 2021 \tia{Oscillations in the stochastic gravitational wave background from sharp features and particle production during inflation} \doij{10.1088/1475-7516/2021/08/030}{J.\ Cosmol.\ Astropart.\ Phys.}{JCAP08}{030}{2021} [\arX{2012.02761}]
\bibitem{Dalianis:2021iig} \au{I}{Dalianis}, \au{G P}{Kodaxis}, \au{I D}{Stamou}, \au{N}{Tetradis} and \au{A}{Tsigkas-Kouvelis} 2021 \tia{Spectrum oscillations from features in the potential of single-field inflation} \doin{10.1103/PhysRevD.104.103510}{Phys.\ Rev.}{D}{104}{103510}{2021} [\arX{2106.02467}]
\bibitem{Boutivas:2022qtl} \au{K}{Boutivas}, \au{I}{Dalianis}, \au{G P}{Kodaxis} and \au{N}{Tetradis} 2022 \tia{The effect of multiple features on the power spectrum in two-field inflation} \doij{10.1088/1475-7516/2022/08/021}{J.\ Cosmol.\ Astropart.\ Phys.}{JCAP08}{021}{2022} [\arX{2203.15605}]
\bibitem{GR}    \au{I S}{Gradshteyn} and \au{I M}{Ryzhik} 2015 \book{Table of Integrals, Series, and Products}{Academic Press}{London}{U.K.}{2007}
\end{thebibliography}
\end{document}